\documentclass[a4paper,11pt]{article}

\usepackage{jheppub} 
\usepackage{lmodern}
\usepackage{tikz}
\usetikzlibrary{decorations.pathmorphing}
\tikzset{snake it/.style={decorate, decoration=snake}}

\usepackage{slashed}
\usepackage[]{mdframed}
\usepackage[T1]{fontenc} 

\newcommand{\be}{\begin{equation}}
\newcommand{\bea}{\begin{eqnarray}}

\newcommand{\ee}{\end{equation}}
\newcommand{\eea}{\end{eqnarray}}

\title{\boldmath Holographic Krylov Complexity for Conformal Quiver Gauge Theories}
\author[a,c]{Ali Fatemiabhari,}
\author[b]{Horatiu Nastase,}
\author[c]{Carlos Nunez}
\author[d]{and Dibakar Roychowdhury}

\affiliation[a]{Institute for Theoretical and Mathematical Physics, Lomonosov Moscow State University, 119991 Moscow, Russia}
\affiliation[b]{Instituto de F\'{\i}sica Te\'orica, UNESP-Universidade Estadual Paulista R. Dr. Bento T. Ferraz 271, Bl. II, Sao Paulo 01140-070, SP, Brazil}
\affiliation[c]{Centre for Quantum Fields and Gravity, Department of Physics, Swansea University, Swansea SA2 8PP, United Kingdom}
\affiliation[d]{Department of Physics, Indian Institute of Technology Roorkee,\\Roorkee 247667, Uttarakhand, India}

\abstract{ We investigate holographic Krylov complexity in fully top-down AdS$_3$ and
AdS$_2$
  supergravity backgrounds dual to two-dimensional linear-quiver SCFTs and
one-dimensional conformal quantum mechanics. In these geometries, the warp
factors, dilaton and other fields depend non-trivially on the `quiver
coordinate' (denoted  by $\eta$ in this paper). This $\eta$-coordinate
encodes the color and flavor data of the dual theories. As a consequence,
a massive probe following a holographic geodesic necessarily moves
simultaneously in the radial AdS direction and along the `quiver direction'. This
produces new contributions to the proper momentum and hence to the rate of
Krylov complexity growth, which is absent in bottom-up AdS models. We show
that the
$\eta$-motion is generically damped, with a time-scale governed by the UV cutoff of the
geodesic
problem, and modifies the early-time evolution of complexity in a
quiver-dependent way. At late times, the $\eta$-dynamics freezes and the
growth
becomes universal, matching pure Poincare AdS predictions. Studying
Abelian and non-Abelian T-dual backgrounds of AdS$_3\times S^3\times T^4$, quivers with localized flavor
groups, and quivers with smeared flavor groups, we quantify how quiver
parameters
shape the operator-spreading dynamics. Our results provide a
systematic characterization of Krylov complexity in top-down
AdS$_3$/AdS$_2$
 duals and reveal a holographic mechanism through which complexity probes
both ultraviolet quiver structure and emergent infrared universality.

}
\begin{document} 
\maketitle
\flushbottom

\section{Introduction and General Idea}
%
%
%
%
%
%
%
%
%
Krylov complexity has recently emerged as a powerful diagnostic of operator growth in quantum many-body systems. In holographic settings, it has been argued that the rate of growth of Krylov complexity is encoded geometrically in the proper momentum of a bulk particle following a radial geodesic, see for example \cite{Caputa:2021sib, Fan:2023ohh, Fan:2024iop, He:2024pox}. This relation was  proposed in JT gravity and in the double-scaled SYK model, see \cite{Rabinovici:2023yex, Xu:2024gfm, Ambrosini:2024sre, Heller:2024ldz} and the reviews \cite{Nandy:2024evd, Rabinovici:2025otw,Baiguera:2025dkc}.

More recently, the authors of \cite{Caputa:2024sux} proposed a concrete formulation in  AdS backgrounds in which the time derivative of the complexity is directly proportional to the proper momentum of a massive probe falling along the AdS radial direction $\bar{\rho}$. 
The precise relation is 
\be
\dot C(t)=-\frac{P_{\bar \rho}}{\epsilon}\;,\label{compmom}
\ee
with $\epsilon$ an UV regulator.  The definition of Krylov complexity was later extended to supersymmetric theories, and 
semiclassical strings in a gravity dual in \cite{Das:2024tnw}.

In \cite{Fatemiabhari:2025cyy},  the notion of proper momentum and proper 
coordinates was generalised  to the case of $AdS_5$ (dual to ${\cal N}=4$ SYM) sliced by $AdS_3$, in order to 
calculate the evolution of holographic Krylov complexity in ${\cal N}=4$ SYM. A further step was taken 
in \cite{Fatemiabhari:2025usn}, where  these notions are extended to the case of 
confining gauge theories with top-down string duals.


The purpose of this paper is to take an essential next step: to analyze holographic Krylov complexity in fully top-down constructions of AdS$_3$ and AdS$_2$
geometries, where the dual field theories are supersymmetric  quiver gauge theories in two and one dimensions. See the papers
\cite{Lozano:2019emq,Lozano:2019jza,Lozano:2019zvg,Lozano:2019ywa,Lozano:2020bxo} for AdS$_3$
and \cite{Lozano:2020txg,Lozano:2020sae,Lozano:2021rmk} for AdS$_2$. We refer to our calculation as Krylov complexity because it is the natural generalisation of that in \cite{Caputa:2024sux} in our case, enriched by warp factors that make our backgrounds dual to conformal quiver gauge theories.

These quivers are highly structured interacting field theories that flow to conformal fixed points in the IR and whose holographic duals are warped AdS backgrounds with nontrivial dependence on an internal $\eta$-coordinate. 
This coordinate encodes the `linear quiver' or `field theory'  direction and is determined by piecewise linear functions $h_4(\eta)$ and $h_8(\eta)$
 that act as rank functions for the gauge groups and flavor content of the dual SCFT (they also encode the Hanany-Witten set up \cite{Hanany:1996ie}). Such solutions preserve four Poincaré supercharges and exhibit an 
$SU(2)$ R-symmetry.

A key observation of this work is that the evolution of holographic Krylov complexity in these top-down geometries cannot be captured by the AdS radial direction alone. Because the warp factors and dilaton depend non-trivially on the quiver coordinate $\eta$, the geodesic of a massive particle necessarily involves simultaneous motion in both $r(t)$ and $\eta(t)$, except for a few and finely-tuned  choices of $h_4,h_8$ rank functions. Consequently, the proper momentum--and therefore the rate of Krylov complexity growth--acquires contributions from the quiver dynamics. This qualitatively novel feature distinguishes top-down constructions from the effective AdS models studied in \cite{Caputa:2024sux} and other papers.

In other words, for bottom-up AdS treatments, the proper momentum and therefore the complexity growth, comes solely from radial infall. However, in the (top-down) supergravity backgrounds dual to 2d and 1d linear-quiver SCFTs, the metric and dilaton depend on the $\eta$-coordinate in such a way that a particle cannot fall purely radially unless the rank functions take very special forms. The equations of motion force a nonzero $\eta(t)$. This means that the particle explores both the renormalization-group $r$-direction and the quiver $\eta$-direction simultaneously.
Physically, this reflects a simple fact: operator growth in a linear quiver CFT inevitably spreads across gauge nodes, because bifundamental matter dynamically couples all nodes. The holographic dual encodes this `spreading' in the warping functions.

We show that the motion in the $\eta$-direction is generically damped and controlled by the UV cutoff scale introduced in the geodesic problem. At early times, the particle explores the quiver direction, and the Krylov complexity receives quiver-dependent corrections. At late times, the 
$\eta$-motion freezes, and the complexity growth becomes indistinguishable from that of pure Poincaré AdS, in agreement with the bottom-up proposal. The UV quiver data influences the initial stages of operator growth, while at long times the dynamics averages over the quiver structure and flows to the one of universal CFTs.

We illustrate these effects across several families of supergravity solutions: Abelian and non-Abelian T-duals of AdS$_3\times S^3\times T^4$, quivers with flavor groups, and quivers with smeared flavor distributions. In each case, we compute the particle trajectories $r(t),\eta(t)$, the proper momentum, and the resulting Krylov complexity. Our findings confirm that top-down holography enriches the structure of complexity growth, providing a more complete dynamical picture of how quiver degrees of freedom enhance  operator spreading.
These results  constitute the first systematic exploration of holographic Krylov complexity in top-down AdS$_3$ and AdS$_2$
 backgrounds encoding nontrivial linear quiver data.

Let us present an outline of the paper. In Section \ref{sec:supergravityback} we write the structure of the supersymmetric AdS$_3$ and AdS$_2$
  backgrounds together with their dual quiver gauge theories. Section \ref{section3} analyses geodesic motion in these warped geometries and shows that radial infall necessarily triggers the motion along the quiver direction. In Section \ref{section3.5} we derive the expression for the proper momentum associated with the combined 
$(r,\eta)$-trajectory and explain its role in computing Krylov complexity. Section \ref{section4} contains our numerical results for several representative families of quiver CFTs and quantum-mechanical models, including Abelian and non-Abelian T-duals, quivers with localised flavor kinks, and quivers with smeared flavor distributions. For each case we determine the dynamics of $r(t)$ and $\eta(t)$, the proper momentum, and the resulting complexity growth, highlighting the influence of quiver data. Finally, Section \ref{sectionconcl} summarizes our findings and discusses future directions.

\section{The supergravity backgrounds and their dual field theories}\label{sec:supergravityback}

Let us start writing the supergravity backgrounds. These solutions to the equations of motion have been constructed in the framework
of holographic duals to conformal field theories in diverse dimensions. Backgrounds of this type have been constructed for cases in which the dual conformal theory has at least an $SO(3)$ R-symmetry and some amount of SUSY is preserved. In this paper we focus our attention on two-dimensional CFTs and one dimensional quantum mechanical systems. We consider the case in which these systems preserve four Poincare SUSYs and have $SU(2)_R$-symmetry. The backgrounds were constructed in  the papers \cite{Lozano:2019emq, Lozano:2019jza, Lozano:2019zvg, Lozano:2019ywa, Lozano:2020bxo, Couzens:2021veb}, for the duals to 2d-SCFTs. For the duals to SUSY quantum mechanics, see the papers \cite{Lozano:2020txg, Lozano:2020sae, Lozano:2021rmk}. there are many other AdS$_3$ and AdS$_2$ backgrounds, but these are not studied here.
\\
\underline{\bf The AdS$_3$ backgrounds and dual SCFTs}
\\
In this work, we only need the Einstein-frame metric (as we study geodesics of particles). Other Ramond and Neveu-Schwarz fields complete these solutions of the supergravity equations of motion. For the case of backgrounds dual to 2d-SCFTs the (massive) IIA 
backgrounds are written in terms of three functions $h_4(\eta),h_8(\eta), u(\eta)$, with their derivatives denoted by $(h_4',h_8',u')$. 
The Einstein frame metric reads
\begin{eqnarray} \label{eq:ads3q}
& & ds_{E}^2= e^{-\frac{\Phi}{2}}\Bigg[ \frac{u}{\sqrt{{h}_4 h_8}}\bigg(\mathrm{d}s^2_{\text{AdS}_3}+\frac{h_8{h}_4 }{4 h_8{h}_4+(u')^2} \mathrm{d}s^2_{\text{S}^2}\bigg)+ \sqrt{\frac{{h}_4}{h_8}} \mathrm{d} s^2_{\text{CY}_2}+ \frac{\sqrt{{h}_4 h_8}}{u} \text{d}\eta^2\Bigg]\nonumber\\
& &e^{-\Phi}= \frac{h_8^{\frac{3}{4}} }{2{h}_4^{\frac{1}{4}}\sqrt{u}}\sqrt{4h_8 {h}_4+(u')^2} .\label{ads3space}  
\end{eqnarray}
The background is a warped product of AdS$_3 \times S^2\times CY_2\times R_\eta$.
The $SU(2)_R$-symmetry of the CFT is  realised by the isometries of the two-sphere, the Calabi-Yau two fold plays an spectator-role and the $\eta$-coordinate is the one on which all fields and warp factors depend. In this paper we choose to write the AdS$_3$ space as
\begin{equation}
ds^2_{AdS_{3}}= e^{-\lambda r}(-dt^2+d\vec{x}^2_{2})+ dr^2,\label{adspoincare3}
\end{equation}
where $\lambda$ is the inverse radius of AdS.

In \cite{Lozano:2019emq} it was shown that the system preserves SUSY. The BPS equations and Bianchi identities for the RR-fields equations are solved if simple ordinary differential equations for $[h_4(\eta),h_8(\eta), u(\eta)]$ are satisfied. These equations read
\begin{eqnarray}
 & &    h_4''= \sum_{j=1}^{P}F_j \delta(\eta-\eta_j), ~~~h_8''= \sum_{j=1}^{P}\tilde{F}_j \delta(\eta-\eta_j).\label{sources}\\
 & & u''=0.\nonumber
\end{eqnarray}
In particular, the right hand side of $h_4''$ and $h_8''$ is associated with localised sources on which the Bianchi identity of the Ramond fields is violated. At those points, the background contain explicit sources. As we briefly discuss below, these localised sources provide the local $SU(F_j)$ gauge groups in the bulk that translate in the presence of global symmetries in the dual CFTs.

The functions $h_4(\eta), h_8(\eta)$ are convex polygonals. They become zero at $\eta=0$ and at $\eta=2\pi (P+1)$--which is the range of the $\eta$-coordinate. At the points $\eta_j$ the polygonal changes slope and a flavour group in the CFT is generated.

The field theory dual to these backgrounds is studied in \cite{Lozano:2019jza, Lozano:2019zvg, Couzens:2021veb}. These papers write quite elaborated quivers, that in the notation customary to two-dimensional SUSY field theory contain: hypers, twisted hypers, Fermi and vector multiplets. There is an associated Hanany-Witten \cite{Hanany:1996ie} set-up consisting of an arrangement of D2, D4, D6 and D8 branes. We do not repeat this lengthy and elaborated analysis here. The conformal field theory describes the IR-fixed point of these  UV-free field theories. Various checks (gauge anomalies cancelation, global anomaly matching, central charge) have been performed, showing agreement between the field theory and the supergravity calculation. For the purposes of this paper, the reader should recall that the UV-quiver field theory is encoded in the functions $h_4, h_8$. These functions act as `rank-functions' for the quiver.

\begin{figure}[h!]
    \centering
    {{\includegraphics[width=10cm]{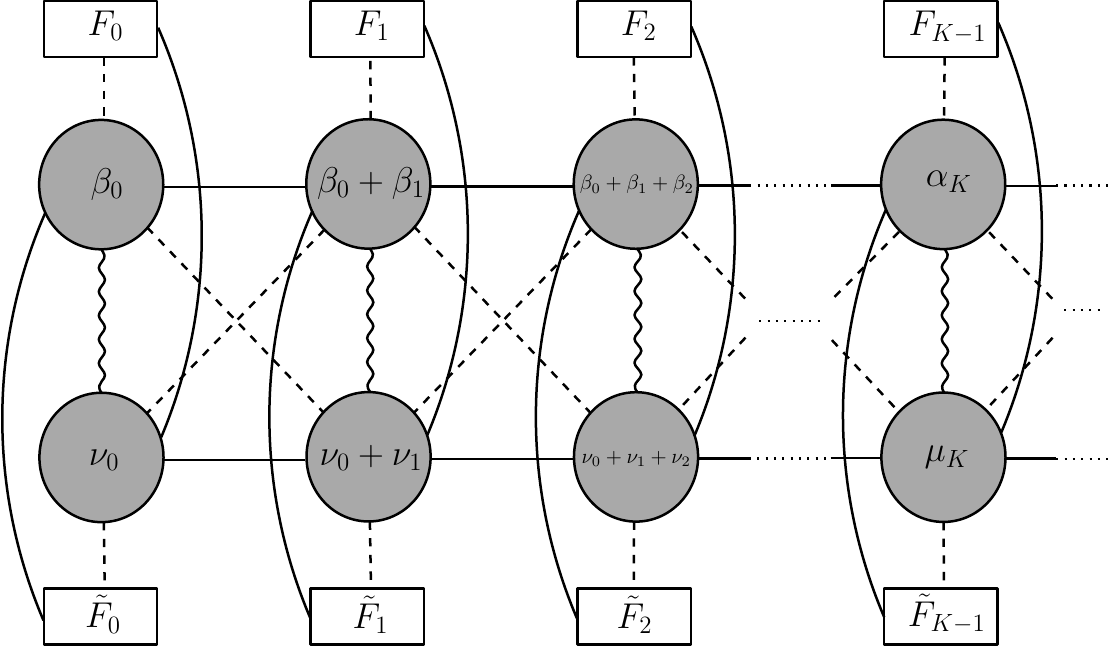} }}%

\caption{ A generic quiver field theory whose IR is dual to the holographic background defined by the functions in \eqref{profileh4sp}-\eqref{profileh8sp}. The solid horizontal black line represents a $(4,4)$ twisted-hypermultiplet connecting two nearest neighbor gauge nodes. Vertically two adjacent gauge nodes are connected by $(0,4)$ hypers represented by wiggly lines and diagonally by $(0,2)$ Fermi multiplets represented by dashed lines. The solid curved lines represent ${\cal N}=(4,4)$ twisted hypers, connecting flavour nodes with opposite gauge nodes.}
\label{figurageneral}

\end{figure}

\begin{figure}[h!]
    \centering
    {{\includegraphics[width=11cm]{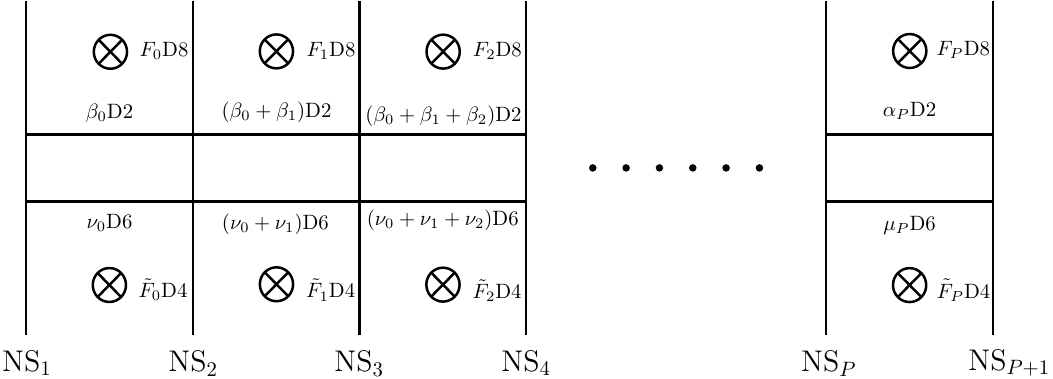} }}%

\caption{Hanany-Witten set-up associated with our generic quiver in figure \ref{figurageneral}. The vertical lines denote NS five branes, horizontal lines D2 and D6 colour branes. The crosses, D4 and D8 flavour branes. }
\label{vvvbb}
\end{figure}

For full details and clear explanations, we refer the reader to the papers \cite{Lozano:2019zvg, Lozano:2019jza, Couzens:2021veb}. Hoping to  add some clarity, we write generic functions $u(\eta)$, $h_4(\eta)$ and $h_8(\eta)$ and the associated two dimensional UV-quiver field theory. Let us define
 \begin{equation} \label{profileh4sp}
 h_4(\eta)\!=\!\!\left\{ \begin{array}{ccrcl}
                       \frac{\beta_0 }{2\pi}
                       \eta & 0\leq \eta\leq 2\pi \\
                 \beta_0+\frac{\beta_1}{2\pi}(\eta-2\pi)      &2\pi \leq \eta \leq 4\pi \\
(\beta_0+\beta_1 )+\frac{\beta_2}{2\pi} (\eta-4\pi) & 4\pi\leq\eta\leq 6\pi\\
                                     (\beta_0+\!\beta_1\!+\!....+\!\beta_{k-1})\! +\! \frac{\beta_k}{2\pi}(\eta-2\pi k) &~~ 2\pi k\leq \eta \leq 2\pi(k+1),\;\;\;\; k:=3,....,P-1\\
                      \alpha_P-  \frac{\alpha_P}{2\pi}(\eta-2\pi P) & 2\pi P\leq \eta \leq 2\pi(P+1).
                                             \end{array}
\right.
\end{equation}
 \begin{equation} \label{profileh8sp}
h_8(\eta)
                    =\left\{ \begin{array}{ccrcl}
                       \frac{\nu_0 }{2\pi}
                       \eta & 0\leq \eta\leq 2\pi \\
                 \nu_0+\frac{\nu_1}{2\pi}(\eta-2\pi)      &2\pi \leq \eta \leq 4\pi \\
(\nu_0+\nu_1 )+\frac{\nu_2}{2\pi} (\eta-4\pi) & 4\pi\leq\eta\leq 6\pi\\
                                     (\nu_0+\nu_1+....+\nu_{k-1}) + \frac{\nu_k}{2\pi}(\eta-2\pi k) &~~ 2\pi k\leq \eta \leq 2\pi(k+1),\;\;\;\; k:=3,....,P-1\\
                      \mu_P-  \frac{\mu_P}{2\pi}(\eta-2\pi P) & 2\pi P\leq \eta \leq 2\pi(P+1).
                                             \end{array}
\right.
\end{equation}
and \begin{equation}
u=\frac{b_0}{2\pi}\eta.\nonumber
\end{equation}
The background in \eqref{eq:ads3q}--plus the Ramond and Neveu-Scharz fields that we did not quote-- for the functions  ${h}_4,h_8, u$ above is dual to the CFT describing the low energy dynamics of a two dimensional quantum field theory encoded by the  quiver in Figure \ref{figurageneral} and the Hanany-Witten set-up of figure \ref{vvvbb}. 

In the following we use the nomenclature for 2d multiplets in \cite{Franco:2015tna, Tong:2014yna}. In these papers general aspects about 2d field theories are clearly explained and suitable to understand the field theory proposal of \cite{Lozano:2019jza, Lozano:2019zvg, Couzens:2021veb}. The papers \cite{Franco:2015tna, Tong:2014yna} are recommended to the interested reader.

In the generic quiver we have two `lines' of gauge groups, made out of $(0,4)$ vector multiplets and $(0,4)$ adjoint hypers. These nodes are horizontally joined by bifundamental matter consisting on $(4,4)$ twisted-hypermultiplets. Vertically, the gauge nodes are joined by $(0,4)$ hypers and diagonally by $(0,2)$ Fermi multiplets. There are flavour symmetries $SU(F_k)\times SU(\tilde{F}_k)$ that are joined to the adjacent gauge nodes by $(0,2)$ Fermi multiplets and with the `opposite' gauge node by $(4,4)$ twisted hypers. This is deduced studying the string quantisation in the Hanany-Witten set-ups of Figure \ref{vvvbb}, see \cite{Couzens:2021veb}.

The ranks of the $SU(\alpha_k)$ gauge nodes (in the upper line of the quiver) are given by $\alpha_k=\sum_{i=0}^{k-1} \beta_i$, being $\beta_i$ the numbers defining the function $h_4(\eta)$. Analogously, the ranks of the $SU(\mu_k)$ gauge nodes in the lower line of the quiver are $\mu_k=\sum_{i=0}^{k-1} \nu_i$, with $\nu_i$ the numbers defining the function $h_8(\eta)$. The flavour nodes have rank $F_k=\nu_{k-1}-\nu_k $
and $\tilde{F}_k=\beta_{k-1}-\beta_k$.
The papers \cite{Couzens:2021veb, Lozano:2019zvg} present detailed explanations for this proposed quivers, showing agreement for certain quantities (matching of 't Hooft anomalies, cancellation of gauge anomalies, central charges) computed both from the field theory and the supergravity pictures.
\\
\underline{\bf The  AdS$_2$ backgrounds and dual conformal quantum mechanics: }
\\
For the AdS$_2$ Einstein frame backgrounds (dual to SUSY quantum mechanics) a very similar logic as in the AdS$_3$ case was used in their construction \cite{Lozano:2020txg, Lozano:2021rmk, Lozano:2020sae}.  In fact, the AdS$_2$ backgrounds in \cite{Lozano:2020txg, Lozano:2020sae, Lozano:2021rmk}
are obtained after a T-duality acting on the massive IIA backgrounds in eq.(\ref{eq:ads3q}), generating Type IIB solutions. The dual conformal quantum mechanics is proposed to be `half' of the SCFT$_2$, in the sense of keeping only the left-sector of it. For details, see \cite{Lozano:2020txg, Lozano:2020sae, Lozano:2021rmk, Filippas:2020qku, Speziali:2019uzn}.

We quote the family of Einstein-frame metrics.
These are written in terms of the AdS$_2$ of radius $\lambda$ that we write in Poincare coordinates as
\begin{equation}
ds^2_{AdS_{2}}= e^{-\lambda r}(-dt^2+d{x}^2)+ dr^2,\label{adspoincare}
\end{equation}
a two sphere, a Calabi-Yau two-fold, the $\eta$-coordinate and a cyclic coordinate $\psi$.
The same functions $[u(\eta), h_4(\eta),h_8(\eta)]$
of the form in eqs.(\ref{profileh4sp})-(\ref{profileh8sp}) solving eq.(\ref{sources}) are used.
\begin{eqnarray}
&\text{d}s^2 &= e^{-\frac{\Phi}{2}}\Bigg[\frac{u}{\sqrt{{h}_4 h_8}} \left( \frac{1}{4}\text{d}s^2_{\text{AdS}_2} + \frac{{h}_4 h_8}{4 {h}_4 h_8 + (u')^2} \text{d}s^2_{\text{S}^2} \right) + \sqrt{\frac{{h}_4}{h_8}} \text{d}s^2_{\text{CY}_2} + \frac{\sqrt{{h}_4 h_8}}{u} (\text{d} \eta^2 +  \text{d} \psi^2 )\,\Bigg] , \nonumber\\
&e^{- 2 \Phi}&= \frac{h_8}{4{h}_4} \Big(4 {h}_4 h_8 + (u')^2 \Big) \, .\label{ads2space}
\end{eqnarray}
Let us close this section with an `executive summary': we presented the metrics of families of AdS$_3$ backgrounds (in massive IIA) and AdS$_2$ in type IIB. Aside from the AdS-factor, a characteristic is the presence of an $\eta$-coordinate, ranging  in $0\leq\eta\leq 2\pi (P+1)$, describing a quiver with $2P$ gauge nodes, bifundamental matter and fundamental matter. The quiver is encoded in the functions $[h_4(\eta),h_8(\eta),u(\eta)]$. In some sense, the $\eta$-coordinate describes the `field theory space'.

In the next section, we study the motion of a particle of mass $m$ that falls under the influence of gravity along the $r$-coordinate. Interestingly, we {\it must} also allow the particle to move in the $\eta$-direction. In some sense, the geodesic explores the `radius/energy' $r$-direction, and also the `field-theory/quiver' $\eta$-direction!

\section{Geodesic motion}\label{section3}

As we anticipated, we study the motion of a particle (that couples to the Einstein frame metric), following a geodesic parametrised by the $t$-coordinate and $r(t), \eta(t)$. 
The induced metric on the particle's worldline is
\begin{equation}
ds_{ind}^2= e^{-\frac{\Phi}{2}}A(\eta) \bigg(  \dot{r}^2- e^{-\lambda r(t)} +\frac{\dot{\eta}^2}{A^2(\eta)} \bigg)dt^2,~~A(\eta)=\frac{u}{\sqrt{h_4 h_8}}.   \label{induced2-3}
\end{equation}
The  main difference between the metrics in eqs.(\ref{ads3space}) and (\ref{ads2space}) is in the dilaton $\Phi(\eta)$.
Hence, the AdS$_2$ and AdS$_3$ cases can be dealt with simultaneously as we do in eq.(\ref{induced2-3}). The factor of one-quarter  difference in both metrics--compare eqs.(\ref{eq:ads3q}) and (\ref{ads2space})-- can be countered by defining in the AdS$_2$ case
\begin{equation}
 r\to 2 r, ~~~\lambda \to \frac{\lambda}{2} ~,~t \rightarrow 2t.\nonumber    
\end{equation}
The action for this probe particle of mass $m$ is
\begin{equation} 
    S=-m \int dt \sqrt{-\det[g_{ind}]}= -m\int dt \sqrt{e^{-\frac{\Phi}{2}} A(\eta)\left( e^{-\lambda r(t)} - \dot{r}^2 -\frac{\dot{\eta}^2}{A(\eta)^2}\right)}.\label{action}
\end{equation}
The equations of motion are
\begin{eqnarray}
& & \frac{d}{dt}\Bigg[ \frac{e^{-\frac{\Phi}{2} } A ~\dot{r}}{L}\Bigg]=\frac{\lambda e^{-\frac{\Phi}{2} -\lambda r}~ A}{{2}L},\label{eqr}
\end{eqnarray}
where $L=\sqrt{e^{-\frac{\Phi}{2}}~A \left( e^{-\lambda r} -\dot{r}^2-\frac{\dot{\eta}^2}{A^2}\right)}$, and
\begin{eqnarray}
& & \frac{d}{dt}\Bigg[ \frac{e^{-\frac{\Phi}{2} }  ~\dot{\eta}}{A ~L}\Bigg]={-\frac{1}{L}\Bigg( \frac{\partial_\eta \left( e^{-\frac{\Phi}{2}} A\right)}{2} \left[e^{-\lambda r} -\dot{r}^2 -\frac{\dot{\eta}^2}{A^2} \right] + e^{-\frac{\Phi}{2}} \frac{\dot{\eta}^2\partial_\eta A}{A^2}    \Bigg)      }.
%
%
\label{eqeta}
\end{eqnarray}

The equations of motion indicate that to have constant $\eta(t)=\eta_0$, (consequently $\dot{\eta}=\ddot{\eta}=0$), we need the derivative
\begin{equation}
\partial_\eta \left(e^{-\frac{\Phi}{2}}A \right)=0.\label{costantetacond}
\end{equation}
Hence, for any background in which the condition in eq.(\ref{costantetacond}) is not satisfied,
a particle moving along the $r$-coordinate will also necessarily move in the $\eta$-direction.

For this geodesic motion, we have a conserved Hamiltonian
\begin{equation}  
H/m=H_0= \frac{e^{-\lambda r -\frac{\Phi(\eta)}{4}} A(\eta)}{\sqrt{A(\eta)\left( e^{-\lambda r(t)} - \dot{r}^2 -\frac{\dot{\eta}^2}{A(\eta)^2}\right)}}.\label{hamiltonian}
\end{equation}
From this Hamiltonian, we find
\begin{equation}
H_0^2\dot{\eta}^2=A^2(\eta)\Big[e^{-\lambda r}H_0^2 - e^{-2\lambda r -\frac{\Phi}{2} }A(\eta) - H_0^2 \dot{r}^2 \Big].\label{etap2}    
\end{equation}
Using eq.(\ref{etap2}), we calculate $\ddot{\eta}(t)$ and replace both $\dot{\eta}(t)$ and $\ddot{\eta}(t)$ in the $r$-equation of motion (\ref{eqr}). We find that the $r$-equation (\ref{eqr}) is satisfied
if
\begin{eqnarray}
& & 2 e^{\lambda r} \ddot{r} +2\lambda e^{\lambda r} \dot{r}^2-\lambda=0, ~\text{implying}\nonumber\\
& & \lambda ~ r(t)= \log\left( c_1 +\frac{\lambda^2(t+c_2)^{{2}}}{4} \right).\label{r(t)}
\end{eqnarray}
Interestingly, this is the same equation that $r(t)$ would have satisfied, had we decoupled the motion from $\eta(t)$!
In other words, the dynamics in $r(t)$ is the same as in pure (unwarped) $AdS$-spaces. This is not so surprising: we are  using holography to analyse the dynamics of an operator in a two-dimensional CFT or in a  conformal quantum mechanical system. Even if the theory has a quiver-like structure, it is still a CFT. What is non-trivial and new is the motion in the $\eta$-coordinate. This can be interpreted as a `motion' along the quiver.

{The integration constants $c_1$ and $c_2$ in eq.(\ref{r(t)}) can be fixed from the initial conditions.
We require that the particle starts its fall in the $r$-coordinate with zero-initial velocity, namely $\dot{r}(t=0)=0$ and with initial position equal to some large UV-cutoff position $r(t=0)=r_{UV}$. These initial conditions yield the values 
\begin{equation}
    c_1=e^{\lambda  r_{UV}}~~ \text{and}~~ c_2 =0.\label{c1c2values}
    \end{equation}
    Substituting these values into eq. \eqref{r(t)}, we find
\begin{align}
    \lambda~ r (t)=\log \left(e^{\lambda ~ r_{UV}}+\frac{\lambda ^2 t^2}{4}\right).\label{solr-t}
\end{align}
In the coordinate we used to write AdS$_{d+1}$, see eqs.~(\ref{adspoincare3}), (\ref{adspoincare}) and considering that $\lambda>0$, the UV of the dual CFT is at $r\to -\infty$. So, $r_{UV}\to -\infty$ is a very negative number. The coordinate $r$ ranges in $(-\infty_{UV}, +\infty_{IR})$. Conversely, if we choose $\lambda<0$, then $r_{UV}$ is very large (positive) number and the coordinate ranges in $(-\infty_{IR}, \infty_{UV})$. Then, eq.(\ref{solr-t}) indicates that at $t=0$, we sit at $r=r_{UV}$, whilst as time goes by,  $r$ grows or decreases depending on the sign of $\lambda$. In fact, the time to reach $r=0$ is
\begin{align}
    t_f=\frac{2}{\lambda}\sqrt{1-e^{\lambda r_{UV}}}>0.
\end{align}
To reach $r=|r_{UV}|$ (we may call this the very far IR), the time elapsed is
\begin{equation}
 t=\frac{\sqrt{8}}{\lambda}\sqrt{\sinh(|\lambda r_{UV}|)}.   
\end{equation}
}
Hence, the problem for the motion along the $r$-coordinate is solved by eq.(\ref{solr-t})--for our chosen initial conditions.
Let us now study the nontrivial motion along the  $\eta$-coordinate. We impose the initial conditions
\begin{equation}
\eta (t=0)=\eta_0 ~~~\text{and}~~~ \dot{\eta}(t=0)=0, 
\end{equation}
 using the Hamiltonian constraint in eq.\eqref{etap2} we find
\begin{align}
\label{9.14}
    e^{\lambda r_{UV}}H^2_0 =A(\eta_0)e^{-\frac{\Phi(\eta_0)}{2}}.
\end{align}

  We need to solve the second order equation of motion for the $\eta$-coordinate  \eqref{eqeta}. This equation is solved if the constraint in eq.(\ref{etap2}) is satisfied and if $r(t)$ is given in eq.(\ref{solr-t}). We  solve for $\eta(t)$ going back to eq.(\ref{etap2}), using   eq.(\ref{solr-t}) for the explicit form of $r(t)$, and integrating. We find
\begin{eqnarray}
& & \frac{H_0^2 \dot{\eta}^2}{16}= \frac{A^2 (\eta) \left(A (\eta_0) e^{-\frac{\Phi (\eta_0) }{2}} -A (\eta) e^{-\frac{\Phi (\eta) }{2}} \right)}{\Big(4 e^{\lambda  r_{UV}}+\lambda ^2 t^2\Big)^2}\longrightarrow\label{eq:Heta}\\
& &  \int \frac{d \eta}{A(\eta)\sqrt{A (\eta_0) e^{-\frac{\Phi (\eta_0) }{2}}-A(\eta)e^{-\frac{\Phi(\eta)}{2}}}}=\frac{4}{H_0}\int \frac{dt}{4 e^{\lambda r_{UV}}+\lambda^2 t^2 }\nonumber\\
   & &=\frac{
   2e^{-\frac{\lambda  r_{UV}}{2}} \arctan \left(\frac{\lambda~ t }{2}  e^{-\frac{\lambda r_{UV}} {2} }\right)}{ \lambda H_0}.\label{eta-t}
\end{eqnarray}

We use eq.\eqref{9.14} to express
\begin{align} \label{eq:etat}
    \int_{\eta_{min}}^{\eta_{max}} \frac{d \eta}{A(\eta)\sqrt{A(\eta_0) e^{-\frac{\Phi(\eta_0)}{2}}-A(\eta)e^{-\frac{\Phi}{2}}}}=
   \frac{
   2e^{-\frac{\lambda  r_{UV}}{2}} \arctan \left(\frac{\lambda~ t }{2}  e^{-\frac{\lambda r_{UV}} {2} }\right)}{ \lambda H_0} .
\end{align}
For different quiver field theories, we have different functions $h_4(\eta),h_8(\eta)$,
which produce different functions $A(\eta)$ and $e^{-\frac{\Phi(\eta)}{2}}$. The integral in eq.(\ref{eq:etat})--equivalently, solving eq.(\ref{eq:Heta})-- gives the the motion along the $\eta$-coordinate as a function of time $\eta(t)$.

Typically, the integral in eq.(\ref{eq:etat}) cannot be solved exactly--the same goes for the ODE in eq.(\ref{eq:Heta}). In the following sections, we perform the integral numerically, for different representative quivers and compute the complexity for these conformal quiver field theories. Before  this we would like to clarify one technical point and then define the {\it proper momentum}.

The technical point goes as follows: from eq.(\ref{eq:Heta}), it might seem that  $\eta(t)=\eta_0$ with $\dot{\eta}=\ddot{\eta}=0$  is a solution. On the other hand, as we observe in eq.(\ref{eqeta}), keeping $\eta(t)$ fixed and $\dot{\eta}(t)=0$ can only be attained if $\partial_\eta\left(e^{-\frac{\Phi{(\eta)}}{2}}A(\eta) \right)=0$.
How is that the constraint into eq.(\ref{eq:Heta}) seems to allow for a solution with $\eta(t)=\eta_0$?

To answer this, let us study the variation of the action eq.\eqref{action} respect to the 
constant $\eta(t)=\eta_0$ solution. That is, perform $\eta(t)=\eta_0+\delta\eta(t)$ where $\delta\eta$ is considered to be small. Keeping terms only up to second order under the square root one finds
\begin{align}
   & S=-m\int dt \nonumber\\&\sqrt{A(\eta_0)e^{-\frac{\Phi(\eta_0)}{2}} \left( e^{-\lambda r(t)} - \dot{r}^2 -\frac{\delta\dot{\eta}^2}{A(\eta_0)^2}\right)+\partial_\eta\Big[A(\eta) e^{-\frac{\Phi(\eta)}{2}}\Big]\Bigg|_{\eta_0}\delta\eta(t)(e^{-\lambda r(t)} - \dot{r}^2 )+ \cdots}\;.\label{expandedt}
\end{align}
This  contains first order terms proportional to the warping factor in the metric. This signals that constant $\eta_0$ {\it cannot} be a generic solution to the EOMs\footnote{If we expand the square-root in eq.(\ref{expandedt}) we find 
\begin{align}
    \delta S=-\frac{m }{2} \frac{e^{\frac{\Phi(\eta_0)}{2}}}{A(\eta_0)}\int dt  \partial_\eta\Big[A(\eta) e^{-\frac{\Phi(\eta)}{2}}\Big]\Bigg|_{\eta_0}\delta\eta(t)+\cdots
\end{align}
where $\delta S=S/H_0 -m \int dt$. The first order in $\delta\eta$ indicates that $\eta(t)=\eta_0$ is not a solution to the equations of motion.}

Let us now study the definition of the {\it proper momentum}, that is an instrumental quantity in the definition of the complexity.

%



\section{Proper Momentum}\label{section3.5}

In this section, we will derive an expression for the proper momentum. For a geodesic that is extended in the $r(t), \eta (t)$ plane, the \emph{proper} momentum is defined as the momentum along the proper radial direction ($\bar{\rho}$), which measures the distance between two points along the geodesic for a fixed time and when all the remaining coordinates are set to be constant. In the metric of AdS$_3$ \eqref{eq:ads3q} and $AdS_2$ (\ref{ads2space}), this corresponds to setting $\Delta t = \Delta x_i=0$ to obtain a metric in $(r,\eta)$ submanifold. Here $\Delta x_i$ collectively denotes all the rest of the coordinates. 

Following the above definition, the proper radial distance on the geodesic connecting two points can be expressed as
\begin{align}
    ds_{E}^2\equiv d\bar \rho^2= e^{-\frac{\Phi}{2}}\Bigg[ \frac{u}{\sqrt{{h}_4 h_8}}dr^2+ \frac{\sqrt{{h}_4 h_8}}{u} \text{d}\eta^2\Bigg]\equiv e^{-\frac{\Phi}{2}} \Bigg[ A(\eta)dr^2+ 1/A(\eta) \text{d}\eta^2\Bigg].
\end{align}

In this section we derive an expression for the proper momentum. This is the quantity needed to compute the complexity, which is the subject of the next section. As our analysis reveals, for the most generic configuration, the Krylov complexity can be obtained through appropriate identification of the proper radial direction and the associated momentum.
Let us choose the convention that $\lambda$ is negative so $r$ moves in the decreasing direction.

Since both $r(t)$ and $\eta(t)$ are functions of time, we can (in principle) invert to obtain $\eta(r)$ or $r(\eta)$. We have 
\begin{align}
    & d\bar \rho = \sqrt{e^{-\frac{\Phi}{2}}A(\eta) \Bigg[ 1+ \frac{1}{A(\eta)^2} \eta'(r)^2\Bigg]} dr \rightarrow \frac{d\bar \rho}{dr}=\frac{\dot {\bar \rho}(t)}{\dot r (t)}=\sqrt{e^{-\frac{\Phi}{2}}A(\eta )\Bigg[ 1+ \frac{1}{A(\eta)^2} \eta'(r)^2\Bigg]},\nonumber \\
    & d\bar \rho =\sqrt{e^{-\frac{\Phi}{2}}A(\eta) \Bigg[ r'(\eta)^2+ \frac{1}{A(\eta)^2} \Bigg]} d\eta \rightarrow \frac{d\bar \rho}{d\eta}=\frac{\dot {\bar \rho}(t)}{\dot \eta (t)}=\sqrt{e^{-\frac{\Phi}{2}}A(\eta )\Bigg[ r'(\eta)^2+ \frac{1}{A(\eta)^2}\Bigg]}= \nonumber\\
    & =r'(\eta)\sqrt{e^{-\frac{\Phi}{2}}A(\eta )\Bigg[ 1+ \frac{1}{A(\eta)^2} \eta'(r)^2\Bigg]} .\label{zaza}
\end{align}
We used the notation $\eta'=\frac{d\eta}{dr}$ and $r'=\frac{dr}{d\eta}$ and that $\eta'(r)=\frac{1}{r'(\eta)}$ in the very last expression of eq.(\ref{zaza}). We define the proper momentum in the $\bar \rho$ direction as
\begin{align}
    & P_{\bar \rho}=\frac{\partial\mathcal{L}}{\partial \dot {\bar \rho}}=\frac{\partial\mathcal{L}}{\partial \dot {r}}\frac{\partial\dot r}{\partial \dot {\bar \rho}}+\frac{\partial\mathcal{L}}{\partial \dot {\eta}}\frac{\partial\dot \eta}{\partial \dot {\bar \rho}}\label{propermomentumprho}\\
    & = \frac{P_r}{\sqrt{e^{-\frac{\Phi}{2}}A(\eta )\Bigg[ 1+ \frac{1}{A(\eta)^2} \eta'(r)^2\Bigg]}}+\frac{P_{\eta}}{\sqrt{e^{-\frac{\Phi}{2}}A(\eta )\Bigg[ r'(\eta)^2+ \frac{1}{A(\eta)^2}\Bigg]}}.\nonumber
\end{align}

Using the action in eq.\eqref{action} and applying the Hamiltonian constraint in eq.\eqref{hamiltonian}, the  proper momentum simplifies considerably. In fact, we have
\begin{align}
  &  P_r=\frac{e^{-\frac{\Phi}{2} } A(\eta ) ~\dot r(t)}{L}= H_0 e^{\lambda r(t)} \dot r(t)\\
   & P_{\eta}=\frac{e^{-\frac{\Phi}{2} } A(\eta ) ~\dot \eta(t)}{A(\eta )^2 L}=\frac{H_0 e^{\lambda r(t)} \dot \eta(t)}{A(\eta(t) )^2}\;,
\end{align}
which together with $L={\sqrt{e^{-\frac{\Phi}{2}}~A \left( e^{-\lambda r} -\dot{r}^2-\frac{\dot{\eta}^2}{A^2}\right)}}$ gives
\begin{align}
  &  P_{\bar \rho}= \frac{P_r+P_{\eta} \eta'(r)}{\sqrt{e^{-\frac{\Phi}{2}}A(\eta )\Bigg[ 1+ \frac{1}{A(\eta)^2} \eta'(r)^2\Bigg]}}=\frac{H_0 e^{\lambda r(t)} \dot r(t)+\frac{H_0 e^{\lambda r(t)} \dot \eta(t)}{A(\eta(t) )^2}\eta'(r)}{\sqrt{e^{-\frac{\Phi}{2}}A(\eta )\Bigg[ 1+ \frac{1}{A(\eta)^2} \eta'(r)^2\Bigg]}} \nonumber\\
  &=\frac{H_0 e^{\lambda r(t)} \dot r(t)\Big(1+\frac{\dot \eta(t)^2}{A(\eta(t) )^2\dot r(t)^2}\Big)}{\sqrt{e^{-\frac{\Phi}{2}}A(\eta )\Bigg[ 1+ \frac{1}{A(\eta)^2} \eta'(r)^2\Bigg]}}= \frac{H_0 e^{\lambda r(t)} \dot r(t)\sqrt{ 1+  \frac{\dot \eta(t)^2}{A(\eta(t) )^2\dot r(t)^2}}}{\sqrt{e^{-\frac{\Phi(\eta(t))}{2}}A(\eta(t) )}}. \label{eq:quivermomentum}
\end{align}

This has a suggestive and useful form. If we impose the {\it constraint}  $\eta=\eta_0$, we see that eq.(\ref{eq:quivermomentum}) gives the result of  \cite{Caputa:2024sux}, rescaled by the prefactor in the AdS-metric. Once we use $r(t)$  in eq.(\ref{solr-t}) and numerically solve for $\eta(t)$ using eq.(\ref{eq:Heta}), the proper momentum
is algebraically easy to obtain. Using this, we compute the complexity, in the next section. Our analysis is aligned with \cite{Caputa:2024sux}, which relates the rate of complexity growth to that with the proper momentum ($P_{\bar{\rho}}$) along the geodesic
\begin{align}
    \partial_t C(t)=-\frac{P_{\bar{\rho}}}{\epsilon}\;,
\end{align}
where $\epsilon$ plays the role of the UV cut-off. This is analogous to eq.(\ref{compmom}), borrowed from \cite{Caputa:2024sux}. The main difference is that our proper momentum  in eq.(\ref{eq:quivermomentum}) includes the influence of the quiver structure.

\section{Numerical Solutions and Krylov Complexity}\label{section4}

With the above formulation completed, the stage is now set to explore various conformal quiver field theories  that are dual to backgrounds with $AdS_3$--see eq.\eqref{eq:ads3q} and $AdS_2$--see eq.(\ref{ads2space})-- factors. In the following, we discuss them under several categories. We start considering linear quivers that are holographically dual to backgrounds obtained from the well-known $AdS_3\times S^3\times K_3$ (or $T^4$) background by the application of T-duality. We continue considering the backgrounds obtained via non-Abelian T-duality from the same seed-solution. After this, we discuss more `genuine' conformal quivers, to close with quivers in which the flavour group is `smeared' to $U(1)^{N_f}$.

Let us summarise briefly the main outcomes of the detailed numerical analysis performed in what follows.
We have infinite families of  SCFTs in dimensions two and one (SUSY quantum mechanics). These field theories in the UV  are described by the quivers discussed in Section \ref{sec:supergravityback}. These quivers flow to a conformal fixed point at low energies, which is what our backgrounds describe.  

A scale is introduced by the UV-cutoff introduced to perform the geodesic calculation. This scale separates `early' from `late' times. The late time behaviour for the complexity is the one found using a bottom-up approach in \cite{Caputa:2024sux}. We observe that the probe particle moves in the $\eta$-direction (it explores the field theory direction of the quiver). The motion is heavily-damped, with a time-scale that is inverse of the UV-cutoff $H_0$.
The early times complexity has the contribution from the $\eta$-motion (and the usual  contribution from the $r$-motion). At late times, the geodesic motion coincides with the one found in the bottom-up approach. Below, we explore the influence of the quiver parameters $(P, \mu,\nu,u_0)$ and the UV-cutoff $H_0$ on the motion, the proper momentum and on the complexity, that is always bigger than in the bottom-up approach.

\subsection{Example I: Abelian T-dual of AdS$_3\times S^3\times T^4$}
The first example we choose is that of a quiver with constant rank function (since it has no kinks, there are no flavours in the quiver CFT). This is characterised by 
\begin{equation}
\label{3.16}
h_4(\eta)=h_{4,0},~~~h_8(\eta)=h_{8,0},~~~u(\eta)=u_0.    
\end{equation}

The corresponding metric in eq.(\ref{ads3space}) is the one we would have obtained by taking AdS$_3\times S^3\times T^4$ and performing a T-duality on a $U(1)$ inside the sphere $S^3$. Similarly, using eqs.(\ref{ads2space}) and (\ref{3.16}) we find a background with an AdS$_2$-factor, that is the same as the one obtained after a second T-duality on AdS$_3$ when we write it as a the fibration over AdS$_2$, see \cite{Lozano:2020txg} for details. 

In this example, the massive particle trajectory is completely characterized by the radial direction $r(t)$ of AdS, while the dynamics along the quiver $\eta$-direction freezes out. 
Indeed, the condition in eq.(\ref{costantetacond}) is satisfied.
This can also be seen using the constraint in the expression for $\dot{\eta}^2$ in eq.(\ref{eq:Heta}), which yields $\eta(t)=\eta_0=$ constant. This is a special case as the warping function of AdS-subspace is a constant, independent of $\eta$ coordinate, so the AdS part is fully decoupled from the $\eta$ direction and the particle trajectory is confined within AdS$_3$. The complexity calculation reproduces the results of \cite{Caputa:2024sux}. 

It might seem obvious that the dynamics is the same as in AdS$_3\times S^3\times T^4$ (as the background we are working with is just the T-dual of it). But what is equivalent under T-duality is the dynamics of a string. Here we are studying the motion of a particle. As we mentioned, particles couple to the Einstein frame. It happens that Einstein and string frame in this case are equivalent, as the dilaton is constant. This is not what happens in our next example.

\subsection{Example II: non-Abelian T-dual of AdS$_3\times S^3\times T^4$}
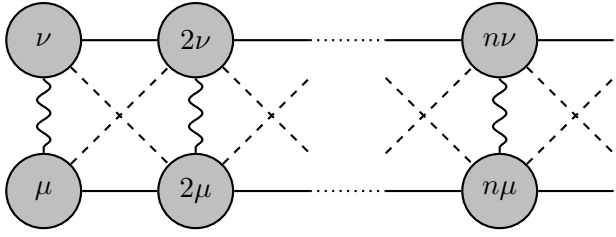
\begin{figure}
\centering
\begin{tikzpicture}[thick]

\tikzstyle{node}=[circle, draw, fill=gray!50, minimum size=28pt, inner sep=2pt]

\node[node] (N4) at (0,2) {$\nu$};
\node[node] (N8) at (0,0) {$\mu$};

\node[node] (2N4) at (2,2) {$2\nu$};
\node[node] (2N8) at (2,0) {$2\mu$};

\node (3N4) at (4,2) {};
\node (3N8) at (4,0) {};

\node[node] (nN4) at (6,2) {$n\nu$};
\node[node] (nN8) at (6,0) {$n\mu$};

\draw[snake it]  (N4) -- (N8);
\draw[snake it] (2N4) -- (2N8);
\draw[snake it] (nN4) -- (nN8);

\draw (N4) -- (2N4);
\draw (N8) -- (2N8);
\draw (2N4) -- (3.5,2);
\draw (2N8) -- (3.5,0);
\draw[dotted] (3.5,2) -- (4.5,2);
\draw[dotted] (3.5,0) -- (4.5,0);
\draw (4.5,2) -- (nN4);
\draw (4.5,0) -- (nN8);
\draw (nN4) -- (7.5,2);
\draw (nN8) -- (7.5,0);

\draw[dashed] (N4) -- (2N8);
\draw[dashed] (N8) -- (2N4);

\draw[dashed] (2N4) -- (3.5,0.5);
\draw[dashed] (2N8) -- (3.5,1.5);
\draw[dashed] (4.5,1.5) -- (nN8);
\draw[dashed] (4.5,0.5) -- (nN4);
\draw[dashed] (nN4) -- (7.5,0.5);
\draw[dashed] (nN8) -- (7.5,1.5);

\end{tikzpicture}

\caption{Quiver associated to the NATD solution.}
\label{fig:quiverNATD}
\end{figure}

We consider the non-Abelian T-dual of AdS$_3\times S^3\times T^4$ (using the left $SU(2)$ inside $SO(4)$ to perform non-Abelian T-duality) \cite{Lozano:2019ywa, Sfetsos:2010uq}. The dual super-conformal quiver does not contain any flavour nodes and is characterised by linearly increasing rank functions of the type 
\begin{equation}
\label{3.21}
h_4(\eta)=\nu \eta,~~~h_8(\eta)=\mu \eta,~~~u(\eta)=u_0\eta.    
\end{equation}
The infinitely long quiver is plotted in Figure \ref{fig:quiverNATD}. This is obviously not a good CFT$_2$: for example, the central charge diverges. How to make sense of this quiver is explained in \cite{Lozano:2016kum, Lozano:2019ywa}. For our purposes, we use this as a simple background where to compute complexity. 

Using eq.\eqref{3.21}, it is easy to compute the functions
\begin{align}
    A(\eta)=\frac{u_0}{\sqrt{\mu \nu}}
\end{align}
and 
\begin{equation} 
e^{-\frac{\Phi (\eta)}{2}}=\left\{ \begin{array}{ccrcl}
&\frac{\sqrt{\frac{(  \mu )^{3/4} \sqrt{4 \eta ^2 \mu  \nu +u_0^2}}{\sqrt[4]{  \nu } \sqrt{  u_0}}}}{\sqrt{2}} & \text{for $AdS_3$} \\
 &\frac{\sqrt{\frac{\sqrt{\mu } \sqrt{4 \eta ^2 \mu  \nu +u_0^2}}{\sqrt{\nu }}}}{\sqrt{2}}&\text{for $AdS_2$}.
\end{array}
\right.
\end{equation}

\begin{figure}
    \centering
    \includegraphics[width=0.5\linewidth]{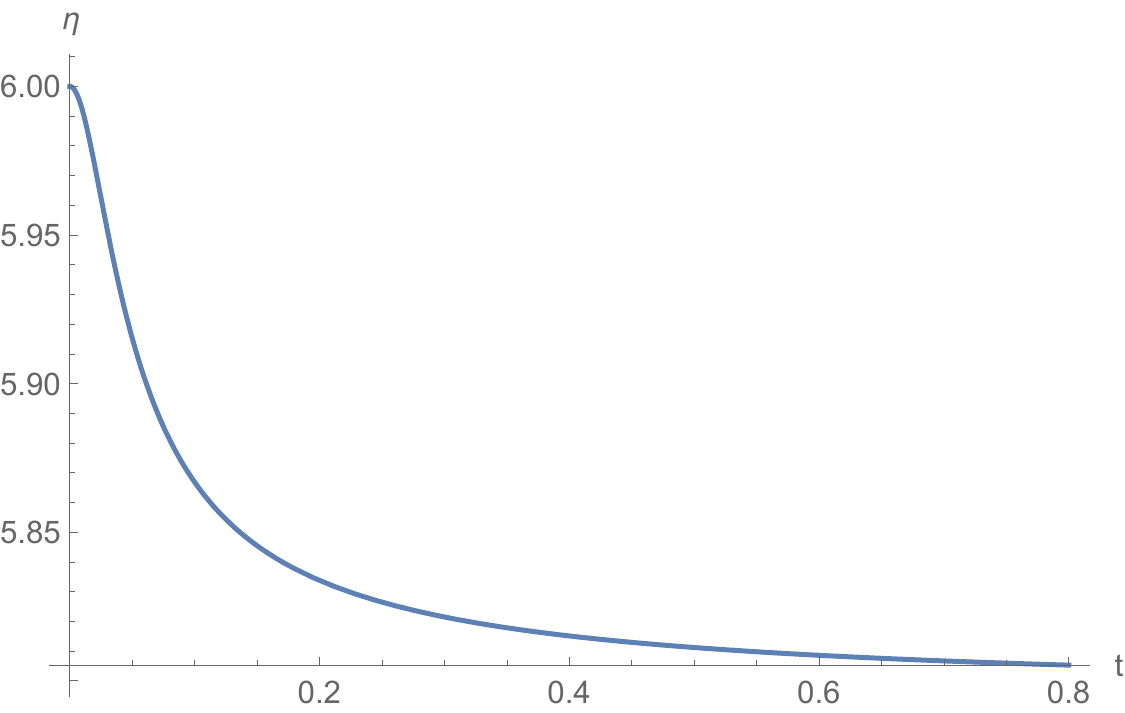}
    \caption{The particle trajectory along the $\eta$ direction for non-Abelian T-dual background. We set $\lambda=\mu=\nu=u_0=1$, $H_0=100$, $\eta_0=6$ and $e^{\lambda r_{UV}}=0.00024$ is fixed by the constraint \eqref{e3.25}.}
    \label{fignatd}
\end{figure}

\begin{figure}
    \centering
    \includegraphics[width=0.45\linewidth]{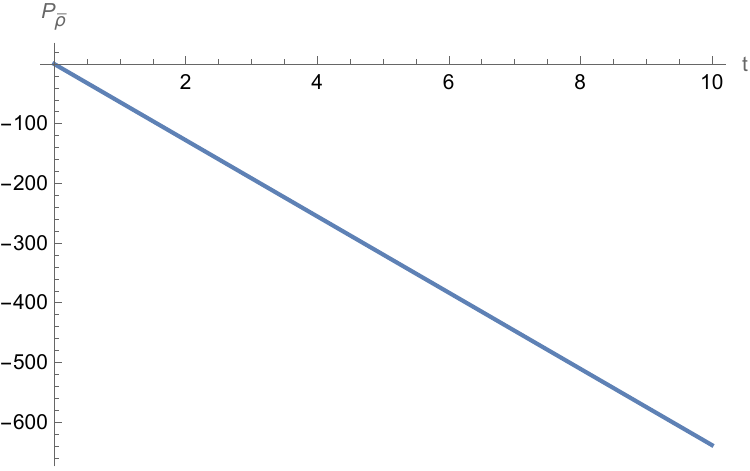}
     \includegraphics[width=0.45\linewidth]{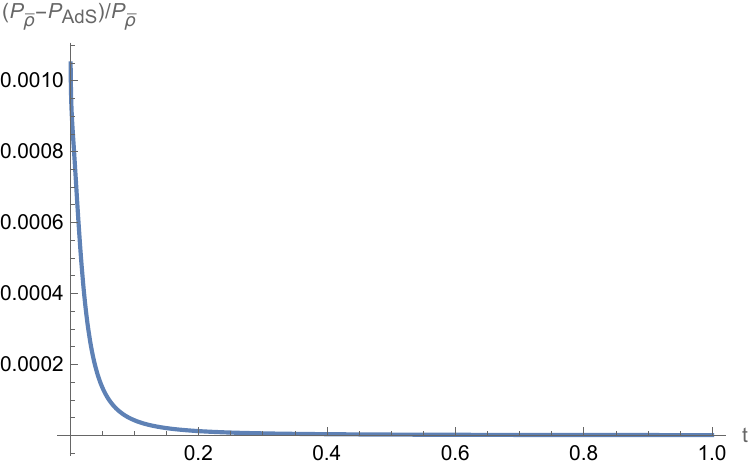}
    \caption{Proper momentum $P_{\bar \rho}$ and its comparison with pure AdS case. $P_{\text{AdS}}$ is obtained by freezing the motion in the $\eta$-direction, as indicated below eq.(\ref{eq:quivermomentum}). We set $\mu=\nu=u_0=1, H_0=100$, and $\eta_0=6$. 
    }
    \label{fig:Pex3}
\end{figure}

We first provide a perturbative expansion for the $\eta(t)$ solution at early times, solving the differential equation \eqref{eq:Heta},
\begin{align}
    \eta(t)\Bigg|_{t\sim0} \sim \eta_0-H_0^2\frac{\eta_0{\mu^{1/8} } \nu ^{5/8} u_0^{5/4}}{\sqrt{2}\left(4 \eta_0^2 \mu  \nu +u_0^2\right)^{5/4}}t^2+O(t)^3,
\end{align}
with $\eta_0=\eta(t=0)$. This suggests that larger $H_0$ leads to faster movement in the $\eta$ direction in early times, while larger $\eta_0, u_0, \mu$ and $\nu$ lead to slower movement. 
For the rate of change of complexity, we have ($\lambda=-2$ is chosen)
\begin{align}
    &\partial_t \mathcal{C}(t)\Bigg|_{t\sim0}    \propto P_{\bar \rho}\Bigg|_{t\sim0}  \sim  -H_0\frac{\sqrt[4]{2}  \sqrt[16]{\mu } \nu ^{5/16}  \sqrt{ \left(16 \eta_0^4 \mu ^2 \nu ^2+u_0^4+9 \eta_0^2 \mu  \nu  u_0^2\right)}}{u_0^{3/8} \left(4 \eta_0^2 \mu  \nu +u_0^2\right)^{9/8}}t + O(t)^2,
\end{align}
which shows the linear growth proportional to $H_0$. Note that the quiver parameters are clearly affecting the rate of change of the complexity.

For numerical studies the integral in eq.(\ref{eta-t}) gets simplified setting  $\mu=\nu=u_0=1$ for simplicity. It can be expressed as
\begin{align}
\label{3.19}
    \int_{\eta_{min}}^{\eta_0} \frac{d \eta}{A(\eta)\sqrt{A(\eta_0)e^{-\frac{\Phi(\eta_0)}{2}}-A(\eta)e^{-\frac{\Phi(\eta)}{2}}}}=\int_{\eta_{min}}^{\eta_0} \frac{d \eta}{\sqrt{e^{-\frac{\Phi (\eta_0)}{2}}-e^{-\frac{\Phi (\eta)}{2}}}},
\end{align}
 Notice that with this choice of parameters the background dilaton $e^{-\frac{\Phi(\eta)}{2}}$ is identical for both AdS$_3$ and AdS$_2$ solutions.

One can numerically solve the integral equation (\ref{3.19}), or equivalently the differential equation \eqref{eq:Heta}, 
\begin{align}
\label{e3.25}
    &\frac{H_0^2 \dot{\eta}^2}{16}= \frac{  \left[e^{-\frac{\Phi (\eta_0)}{2}} -  e^{-\frac{\Phi (\eta)}{2}} \right]}{\Big( \lambda^2 t^2 +4 e^{\lambda r_{UV}}\Big)^2}~;~e^{\lambda r_{UV}}=\frac{A(\eta_0)e^{-\frac{\Phi (\eta_0)}{2}}}{H^2_0}\;,\end{align}
with 
\begin{align} e^{-\frac{\Phi (\eta)}{2}} =\frac{\sqrt[4]{4 \eta ^2+1}}{\sqrt{2}}~;~A(\eta)=1. \nonumber
\end{align}

The full numerical solution of $\eta(t)$ for certain choices of parameters is shown in Figure  \ref{fignatd} and the corresponding proper momentum and comparison with pure $AdS_3$ case is given in Figure \ref{fig:Pex3}.

As can be seen from both  Figures \ref{fignatd} and  \ref{fig:Pex3}, the particle moves in a direction of decreasing ranks of the quiver (towards $\eta\to 0$), and the momentum along the quiver coordinate (the $\eta$-coordinate) freezes out after a short period of time. The range of motion in the $\eta$ direction is not considerably affected by the value of UV cut-off, $H_0$. As we discuss below, $H_0$ introduces a scale, that is inverse with the time-scale in which the $\eta$-motion is damped. The difference with the pure $AdS_3$ calculation is negligible at late times. Interestingly the motion along the quiver $\eta$-direction, affects the complexity only at early times, as depicted in the right panel of Figure \ref{fig:Pex3}. Note that the complexity is larger than in the pure AdS case. 

At late times, the momentum (and hence the Krylov complexity) receives a contribution only from the motion along AdS. This is further depicted in Figure \ref{fig:Pex3}, which reveals a linear growth of complexity with time, which is in agreement with the analysis of the Poincare AdS in \cite{Caputa:2024sux}. Let us study in more detail the parametric dependence.

\subsubsection{Study of quiver parameters}
\begin{figure}
    \centering
     \includegraphics[width=0.6\linewidth]{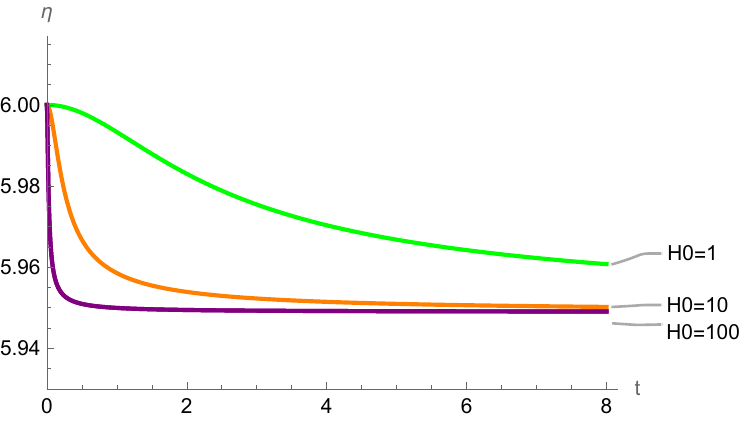}
    \caption{The trajectory in the $\eta$ direction for various $H_0$ choices and $u_0=\mu=\nu=1$.  }
    \label{fig:H0}
\end{figure}

Now, we study the effect of varying different parameters in the problem on the trajectory of the particle in the $\eta$ direction. 
Figure \ref{fig:H0} shows the trajectory in example II for different $H_0$ choices. It is observed that the range of the motion in the $\eta$ direction is not considerably affected by the value of $H_0$. But the rate of approach to the asymptotic value of $\eta_{min}\equiv\eta(t\to\infty)$ is much faster for larger values of $H_0$.

\begin{figure}
    \centering
     \hspace{0.45cm}\includegraphics[width=0.63\linewidth]{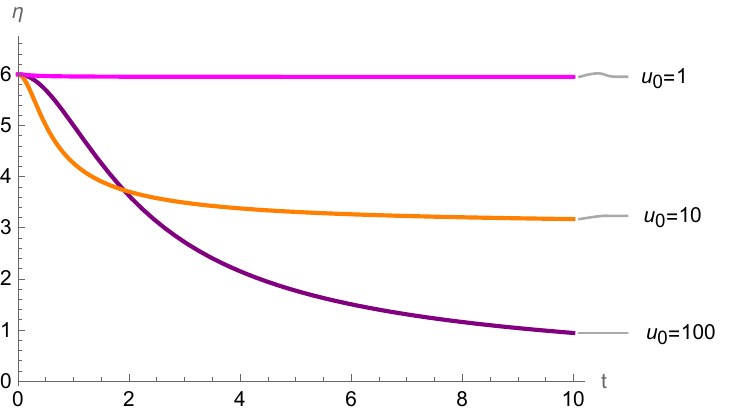}
     \includegraphics[width=0.63\linewidth]{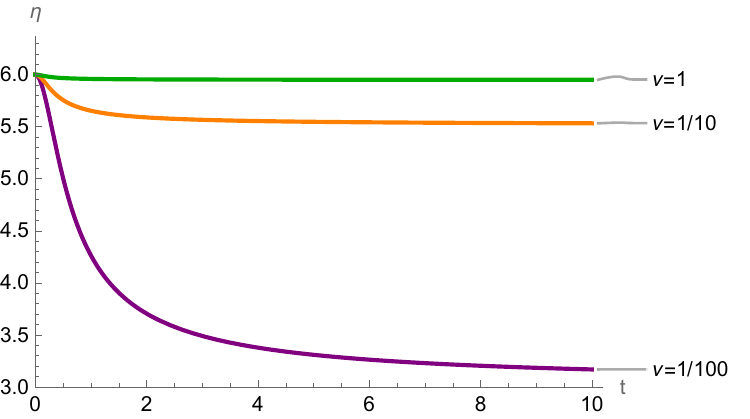}
    \caption{The trajectory in the $\eta$ direction for various $u_0$ choices, $\mu=\nu=1$ (top) and various $\nu$ choices, $\mu=u_0=1$ (bottom) with $H_0=10$.
    }
    \label{fig:u0}
\end{figure}

The top and bottom panels in Figure \ref{fig:u0} depict the trajectory for different choices of $u_0$ and $\nu$, respectively. One can see that increasing $u_0$ enhances the range of motion in the $\eta$ direction, such that for larger values of $u_0$ the particle can approach the origin of the quiver direction at $\eta=0$. The parameter $\nu$ has an inverse effect, and only for smaller values of it, the range of the motion is considerable. One should keep in mind that in the dual field theory, $\nu$ is related to the number of colours, so it should be kept large to be in a reliable gravity dual. This freezes the motion in the $\eta$ direction considerably. The effect of the $\mu$ parameter is similar. One should be wary of taking $(\nu,\mu)$ smaller than one, as we emphasised above. In fact, the supergravity approximation is reliable in the limit of long quivers ($P$ large) and large ranks for each node.

Let us now study a more realistic field theory.

\subsection{Example III: Quivers with a kink}

As our third example, we consider the effect that the  flavour nodes of the  quiver theory have on the complexity. We consider simple examples where the dual supergravity solutions are characterised by  rank functions of the form
\begin{equation}
h_4(\eta)=h_{4,0},~~~u(\eta)=u_0,   
\end{equation}
and 
\begin{equation} \label{profilegeneric}
h_8(\eta)
=\left\{ \begin{array}{ccrcl}
\frac{\nu_0 }{2\pi}\eta & 0\leq \eta\leq 2\pi \\
\mu_k+ \frac{\nu_k}{2\pi}(\eta-2\pi k) &~~ 2\pi k\leq \eta \leq 2\pi(k+1),\;\;\;\; k:=1,....,P-1\\
\mu_P-  \frac{\mu_P}{2\pi}(\eta-2\pi P) & 2\pi P\leq \eta \leq 2\pi(P+1)\;,
\end{array}
\right.
\end{equation}
where only the $h_8(\eta)$ function is introducing the flavour-nodes effects on the geometry.

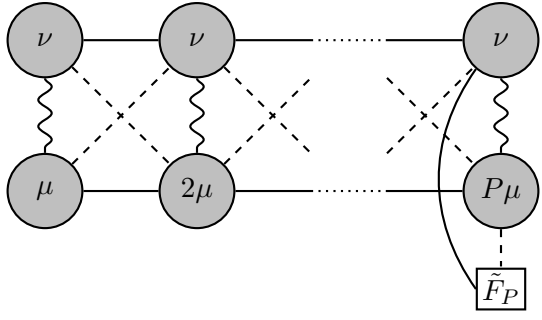
\begin{figure}
\centering
\begin{tikzpicture}[thick]

\tikzstyle{node}=[circle, draw, fill=gray!50, minimum size=28pt, inner sep=2pt]
\tikzstyle{flavor}=[rectangle, draw, minimum size=15pt, inner sep=2pt]

\node[node] (N4) at (0,2) {$\nu$};
\node[node] (N8) at (0,0) {$\mu$};

\node[node] (2N4) at (2,2) {$\nu$};
\node[node] (2N8) at (2,0) {$2\mu$};

\node (3N4) at (4,2) {};
\node (3N8) at (4,0) {};

\node[node] (nN4) at (6,2) {$\nu$};
\node[node] (nN8) at (6,0) {$P\mu$};

\node[flavor] (F8) at (6,-1.3) {$\tilde{F}_{P}$};

\draw[snake it]  (N4) -- (N8);
\draw[snake it] (2N4) -- (2N8);
\draw[snake it] (nN4) -- (nN8);

\draw (N4) -- (2N4);
\draw (N8) -- (2N8);
\draw (2N4) -- (3.5,2);
\draw (2N8) -- (3.5,0);
\draw[dotted] (3.5,2) -- (4.5,2);
\draw[dotted] (3.5,0) -- (4.5,0);
\draw (4.5,2) -- (nN4);
\draw (4.5,0) -- (nN8);

\draw[dashed] (N4) -- (2N8);
\draw[dashed] (N8) -- (2N4);

\draw[dashed] (2N4) -- (3.5,0.5);
\draw[dashed] (2N8) -- (3.5,1.5);
\draw[dashed] (4.5,1.5) -- (nN8);
\draw[dashed] (4.5,0.5) -- (nN4);
\draw[dashed]  (nN8) -- (6,-1);
\draw  (nN4) to [bend right=35] (5.67,-1.3);

\end{tikzpicture}

\caption{Quiver associated to the example III .}
\label{fig:quiverKink}
\end{figure}

The simplest situation is adding one stack of flavour branes at a certain point, namely $\eta= 2\pi P$, along the quiver axis. These ``single kink'' quivers  are given by the following rank function
\begin{equation} \label{profileh8sp1}
h_8(\eta)
=\left\{ \begin{array}{ccrcl}
\frac{\mu }{2\pi}\eta, & 0\leq \eta \leq 2\pi P \\
\mu P-  \frac{\mu P}{2\pi}(\eta-2\pi P), &~~ 2\pi P\leq \eta \leq 2\pi(P+1).
\end{array}
\right.
\end{equation}
The quiver diagram  is provided in Figure \ref{fig:quiverKink}.

If at initial time $t=0$, we put the particle at the position $\eta(t)=\eta_0$, in the $0<\eta_0< 2\pi P$, we find 
\begin{align}
    A(\eta)=\frac{\sqrt{2 \pi }}{\sqrt{\eta  \mu }}
\end{align}
and
\begin{equation} 
A(\eta) e^{-\frac{\Phi(\eta)}{2}}=\left\{ \begin{array}{ccrcl}
&\frac{(\eta \mu)^{1/8}}{(2 \pi )^{1/8}} & \text{for $AdS_3$} \\
 &1&\text{for $AdS_2$}\;,
\end{array}
\right.
\end{equation}
where we set $h_{4,0}=u_0=1$. 

In what follows, we focus our analysis on AdS$_3$ only, as for AdS$_2$ the particle dynamics is decoupled from the quiver and is solely confined to AdS-spacetime. 
This can be seen both from eqs.(\ref{costantetacond}) and \eqref{eq:Heta}, which yield $\dot{\eta}=0$. With our focus now on AdS$_3$, the corresponding integral \eqref{eta-t} turns out to be
\begin{align}
\label{e3.23}
    \int_{\eta_{min}}^{\eta_0} \frac{d \eta}{A(\eta)\sqrt{A(\eta_0) e^{-\frac{\Phi(\eta_0)}{2}}-A(\eta)e^{-\frac{\Phi(\eta)}{2}}}}=\int_{\eta_{min}}^{\eta_0} \frac{d \eta}{(2 \pi)^{7/16}}\frac{\sqrt{\eta \mu}}{\sqrt{(\eta_0 \mu)^{1/8}-(\eta \mu)^{1/8}}}.
\end{align}

\begin{figure}
    \centering
    \includegraphics[width=0.45\linewidth]{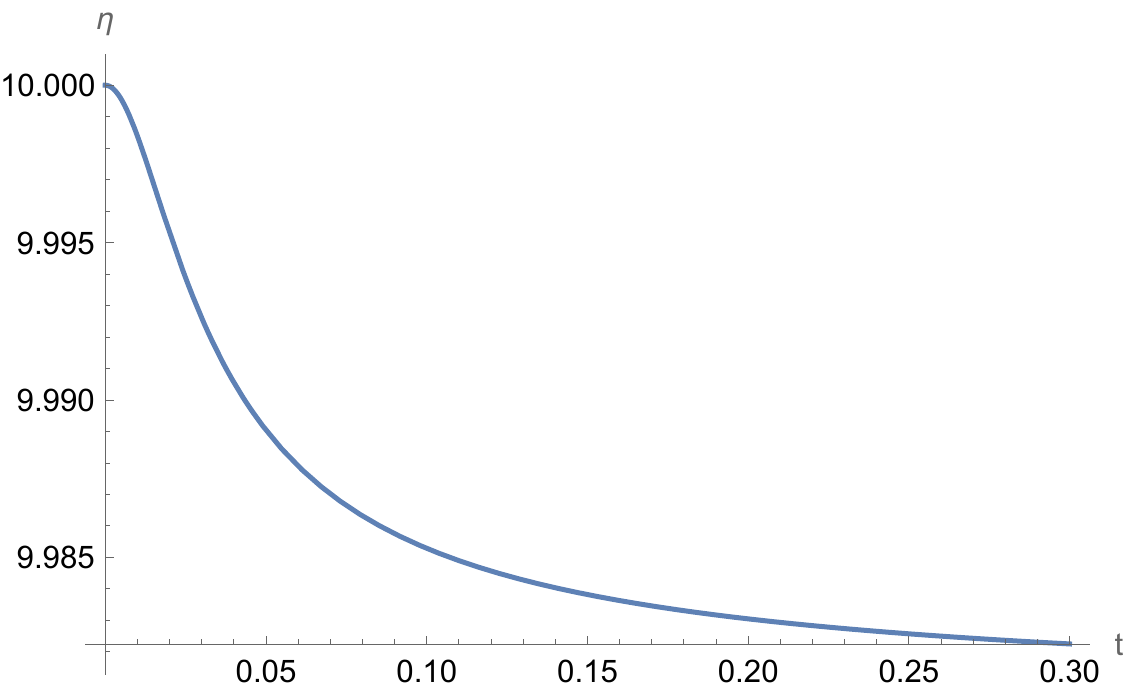}
    \caption{The particle trajectory along the $\eta$ direction for the $AdS_3$ case of  Example III. $\eta_0=10$, $\mu=1$, $H_0=100$ is chosen and $e^{\lambda r_{UV}}=0.0001$ is fixed by the constraint.}
    \label{figex2}
\end{figure}

\begin{figure}
    \centering
    \includegraphics[width=0.45\linewidth]{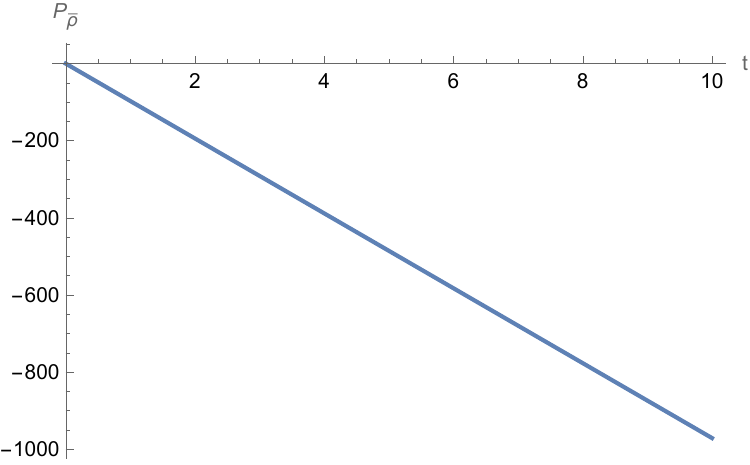}
    \includegraphics[width=0.45\linewidth]{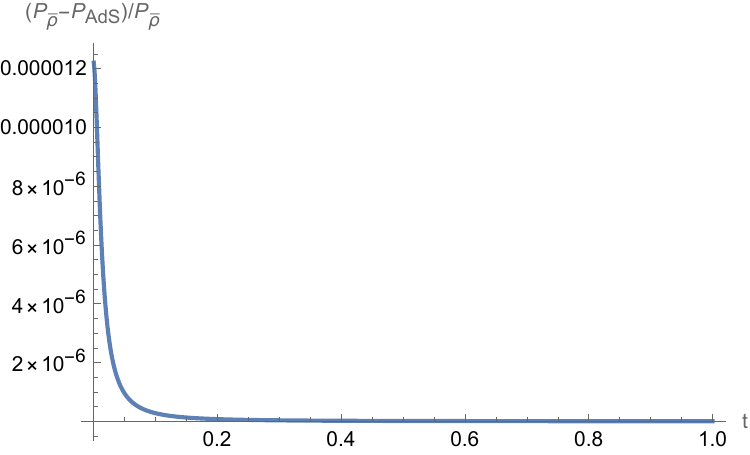}
    \caption{Proper momentum $P_{\bar \rho}$ for the $AdS_3$ case and its comparison with pure $AdS_3$. We set $\eta_0=10$, $\mu=1$, $H_0=100$. } 
    \label{fig:Pex2}
\end{figure}

\begin{figure}
    \centering
    \includegraphics[width=0.45\linewidth]{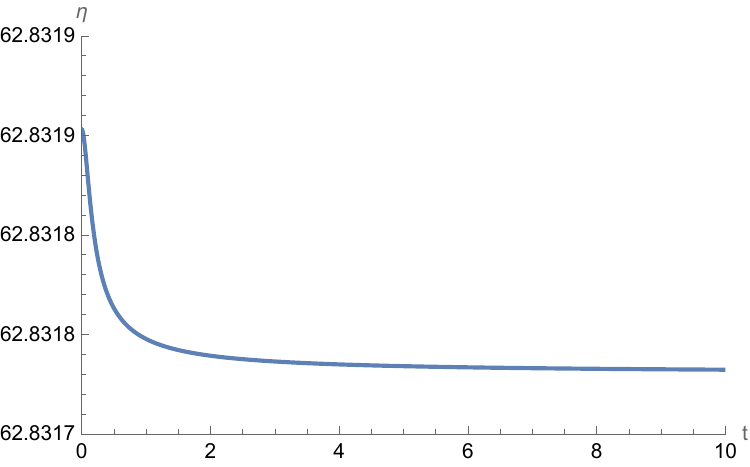}
    \caption{The particle trajectory along the $\eta$ direction for Example III. We set $\eta_0=2 \pi P$, $P=100$, $\mu=1$, $H_0=100$. } 
    \label{figex2a}
\end{figure}

Like the previous example, we first provide a perturbative expansion for the $\eta(t)$, in the range $0<\eta<2\pi P$, at early times as
\begin{align}
    \eta(t)\Bigg|_{t\sim0} \sim \eta_0-H_0^2\frac{\pi ^{9/8}}{8\ 2^{7/8} \eta_0^{17/8} \mu^{9/8}}t^2+O(t)^3.
\end{align}
Here $\eta_0=\eta(t=0)$. This suggests that larger $H_0$ leads to faster movement in the $\eta$ direction in early times, while larger $\eta_0$ and $\mu$ lead to slower movement. 
For the rate of change of complexity, we have
\begin{align}
    &\partial_t \mathcal{C}(t)\Bigg|_{t\sim0}    \propto P_{\bar \rho}\Bigg|_{t\sim0}  \sim  -H_0\frac{\sqrt[16]{\pi } \sqrt{\frac{\pi }{\eta_0^3 \mu }+128}}{8 \times 2^{7/16} \sqrt[16]{\eta_0 \mu }}t + O(t)^2,
\end{align}
which shows the linear growth proportional to $H_0$. Note that the quiver parameters are clearly affecting the rate of change of the complexity. 

Like before, for the full trajectory, the integral \eqref{eta-t} (or the differential eq.\eqref{eq:Heta}) has to be solved numerically, which yields a trajectory shown in Figure \ref{figex2}. As one can see, the particle's position in $\eta$ direction is decreasing, 
and its motion is restricted in a narrow region of $\eta_{min}\leq \eta \leq \eta_0$, where 
$\eta_{min}$ is the $\eta$-position
to which the particle approaches asymptotically. In other words, the particle tends to move towards regions associated with smaller rank function values in the dual gauge theory. 
This is a behaviour similar to that observed in example II. The corresponding proper momentum is provided in Figure \ref{fig:Pex2}.

As a further illustration of our previous calculation, we place the particle at the location of the flavour branes, namely at $\eta_0= 2\pi P$, where we choose $P\gg 1$. We separately perform our analysis for the two intervals. For the first interval $\eta_{min}\leq \eta \leq 2\pi P$, one gets a trajectory as shown in Fig.\ref{figex2a}, which shows that the particle moves towards the decreasing direction of the rank function and finally stops at some $\eta=\eta_{min}$.

For the second interval $2\pi P \leq \eta \leq 2\pi (P+1)$, we find
\begin{align}
    A(\eta)=\frac{\sqrt{2 \pi }}{\sqrt{\mu P (\eta -2 \pi  (P-1))}}\;,
\end{align}
together with the combination
\begin{equation} 
A(\eta) e^{-\frac{\Phi (\eta)}{2}}=\left\{ \begin{array}{ccrcl}
&\frac{(\mu  P(\eta -2 \pi  (P-1)))^{1/8}}{\sqrt[8]{2 \pi }} & \text{for $AdS_3$} \\
 &1&\text{for $AdS_2$}\;,
\end{array}
\right.
\end{equation}
where we set $h_{4,0}=u_0=1$.

\begin{figure}
    \centering
    \includegraphics[width=0.45\linewidth]{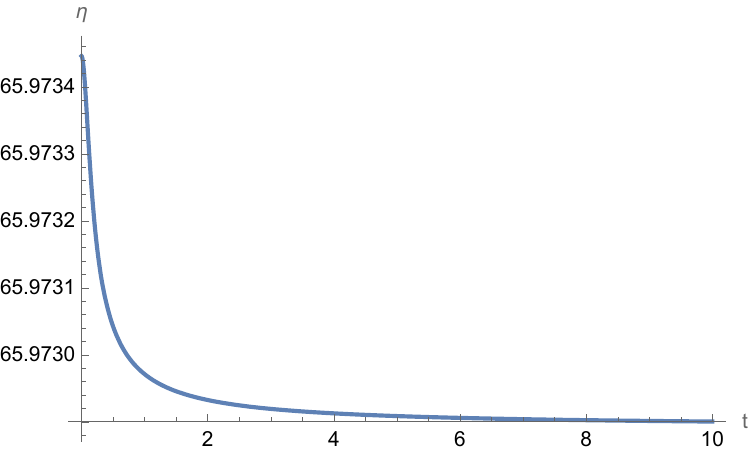}
    \includegraphics[width=0.45\linewidth]{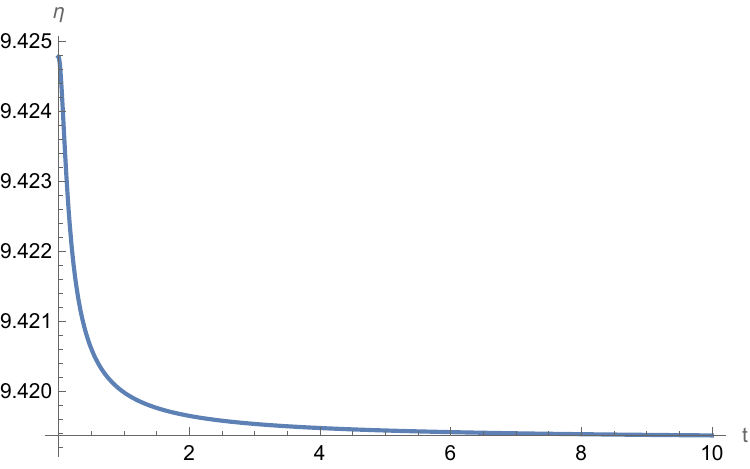}
    \caption{The particle trajectory along the $\eta$ direction for the $AdS_3$ case of  Example III, $P=10$ (left) and $P=1$ (right). $\eta_0=2\pi(P+1/2)$, $\mu=1$ and $H_0=10$ is chosen. Note that the case $P=1$ is done only as a sample calculation. The supergravity approximation is valid for large $P$.}
    \label{fig:quiverP}
\end{figure}

Like before, we repeat our analysis for the second interval, where we take the initial condition as $\eta_0=2\pi (P+1/2)$ and set $\mu =1$. This yields a configuration as shown in the left panel of the Figure \ref{fig:quiverP}. This clearly shows that the particle always moves towards the decreasing direction of the $\eta$ coordinate. 
Notice that, with the flavour brane located near the end of the space ($\eta_0 \sim 2\pi P\gg 1$), the particle always prefers to move in the first interval rather than the second one. 

The right panel of the Figure \ref{fig:quiverP} shows the effect of the length of the quiver $P$ on the trajectory of the particle. Lower values of $P$ leads to larger range of motion in the $\eta$ direction. But one needs to keep in mind that the supergravity solution is a reliable dual to the SCFT when  $P$ is large.

As our next example, we exchange rank functions $h_4\leftrightarrow h_8$ and study their effect on the complexity. For the first interval $0\leq \eta \leq 2\pi P$, with $\eta_0=\eta(t=0)$, we choose $h_8=1=u$ and $h_4=\frac{\mu }{2\pi}\eta$. This yields the functions
\begin{align}
    A(\eta)=\frac{\sqrt{2 \pi }}{\sqrt{\eta  \mu }}
\end{align}
and
\begin{equation} 
\label{e3.29}
A(\eta) e^{-\frac{\Phi}{2}}=\left\{ \begin{array}{ccrcl}
&\frac{(2 \pi )^{3/8}}{(\eta  \mu )^{3/8}} & \text{for $AdS_3$} \\
 &\frac{\sqrt{2 \pi }}{\sqrt{\eta  \mu }}&\text{for $AdS_2$}.
\end{array}
\right.
\end{equation}
Notice that while the function $A(\eta)$ remains unchanged, the dilaton of the background is sensitive to the swapping of the rank functions, which is also evident from eq. \eqref{eq:ads3q}. In other words, the complexity would be different under the exchange of functions $h_4$ and $h_8$.

\begin{figure}
    \centering
    \includegraphics[width=0.5\linewidth]{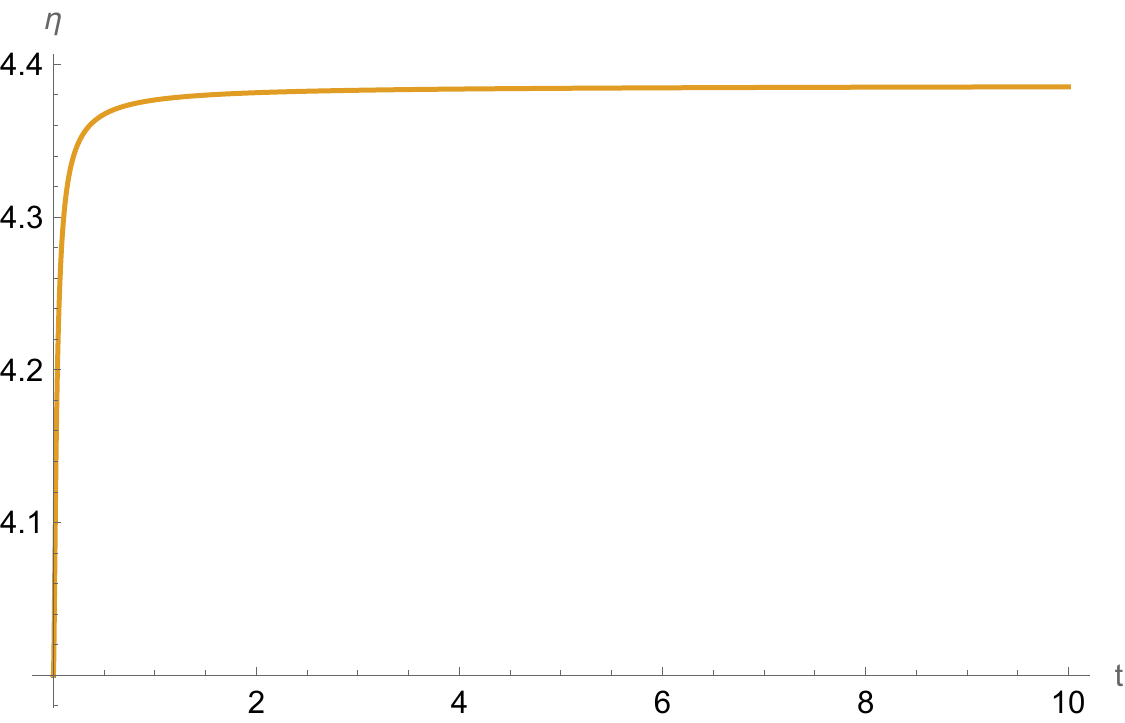}
    \caption{The particle trajectory along the $\eta$ direction for non trivial $h_4$ in Example III with $\eta_0=4$, $\mu=1$, $H_0=100$. 
    }
    \label{exchanged}
\end{figure}

\begin{figure}
    \centering
    \includegraphics[width=0.45\linewidth]{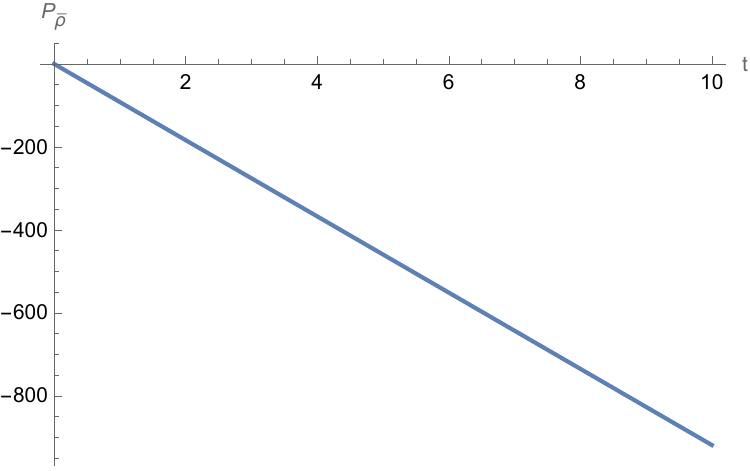}
    \includegraphics[width=0.45\linewidth]{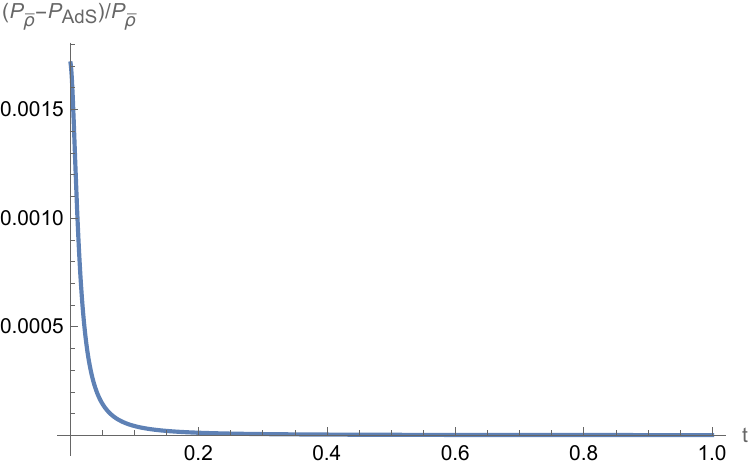}
    \caption{Proper momentum $P_{\bar \rho}$ for non trivial $h_4$ and its comparison with pure AdS case. We set $\eta_0=4$, $\mu=1$, $H_0=100$.}
    \label{fig:exchangedM}
\end{figure}

This is precisely reflected once we proceed with the numeric calculation and solve the differential eq. \eqref{eq:Heta}.  
Notice that in eq.\eqref{eq:Heta}, its right-hand side is positive definite if one satisfies $\eta (t)>\eta_0$. 
This produces a characteristically opposite behaviour compared to the previous cases; see Figure \ref{exchanged}. In other words, unlike previous cases, the particle motion tends towards a direction of increasing value of the rank function.  The corresponding proper momentum is given in Figure \ref{fig:exchangedM}.

\subsection{Example IV: Quivers with smeared flavours}

As our last example, we consider quivers that are characterised by the following rank functions
\begin{equation}
\label{e3.27}
h_4(\eta)=h_{4,0}\sin\left( \frac{\pi \eta}{P+1}\right),~~~h_8(\eta)=h_{8,0}\sin\left( \frac{\pi \eta}{P+1}\right),~~~u(\eta)=u_0   .
\end{equation}
This is a very peculiar situation. In fact the functions $h_4, h_8$ do {\it not} satisfy the equation of motion \ref{sources}.
One may wonder, what is the meaning of these $h_4,h_8$ functions. This choice indicates a smeared distribution of flavour branes. In fact, as explained in the papers \cite{Lozano:2019zvg, Lozano:2019emq}, the Bianchi identity for the Ramond forms $F_0$ and $F_4$ are violated wherever $h_8''$ and $h_4''$ are nonzero. In this case, we find a continuous distribution of sources (in a different system, an example in which a continuous distribution of sources made physical sense is \cite{Filippas:2019puw}). The $\eta$-direction ranges in $[0, (P+1)\pi]$. We should see this example as a sample calculation.

Given \eqref{e3.27}, we have
\begin{align}
    A(\eta)=\frac{1}{\sin\left(\frac{\pi  \eta }{P+1}\right)}
\end{align}
and
\begin{equation} 
A(\eta)e^{-\frac{\Phi (\eta)}{2}}=\left\{ \begin{array}{ccrcl}
&\sin ^{-1/4}\left(\frac{\pi  \eta }{P+1}\right) & \text{for $AdS_3$} \\
 &\sin ^{-1/2}\left(\frac{\pi  \eta }{P+1}\right)&\text{for $AdS_2$}\;,
\end{array}
\right.
\end{equation}
where we set $h_{4,0}=h_{8,0}=u_0=1$ for simplicity.

\begin{figure}
    \centering
    \includegraphics[width=0.8\linewidth]{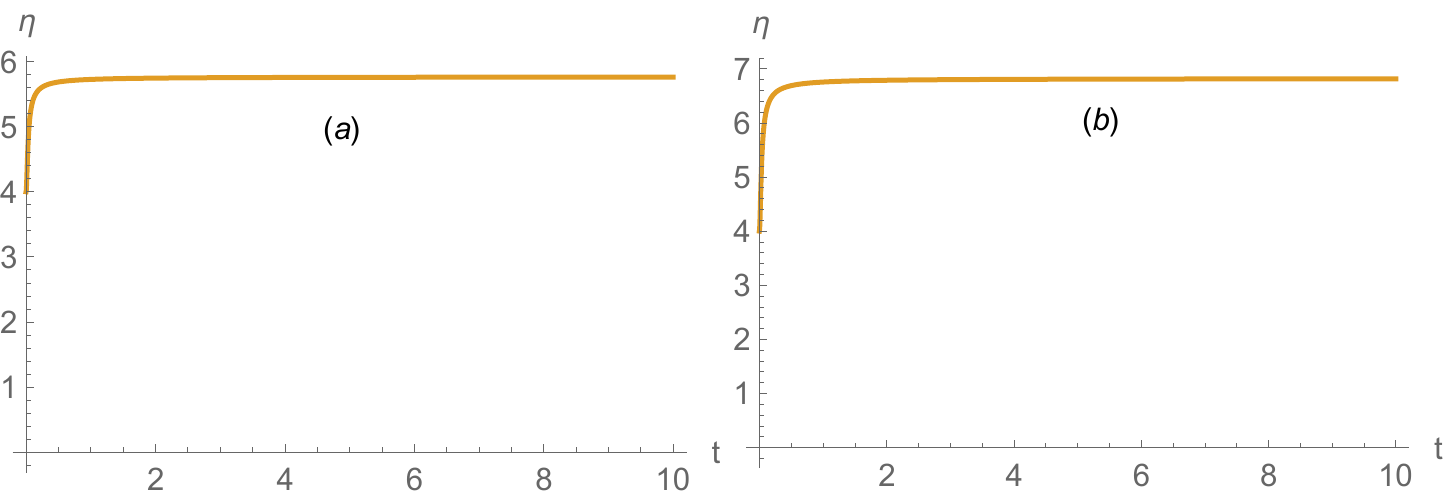}
    \caption{The particle trajectory along the $\eta$ direction for a smeared quiver in Example IV. Fig.(a) is the plot for $AdS_3$. We set $H_0=100$, $\eta_0=4$, $P=50$ and $e^{\lambda r_{UV}}=0.00014$ is fixed by the constraint \eqref{e3.25}. Fig.(b) is the plot for $AdS_2$. We set $H_0=100$, $\eta_0=4$, $P=50$ and $e^{\lambda r_{UV}}=0.0002$ is fixed by the constraint \eqref{e3.25}.}
    \label{figsmeared}
\end{figure}

\begin{figure}
    \centering
    \includegraphics[width=0.45\linewidth]{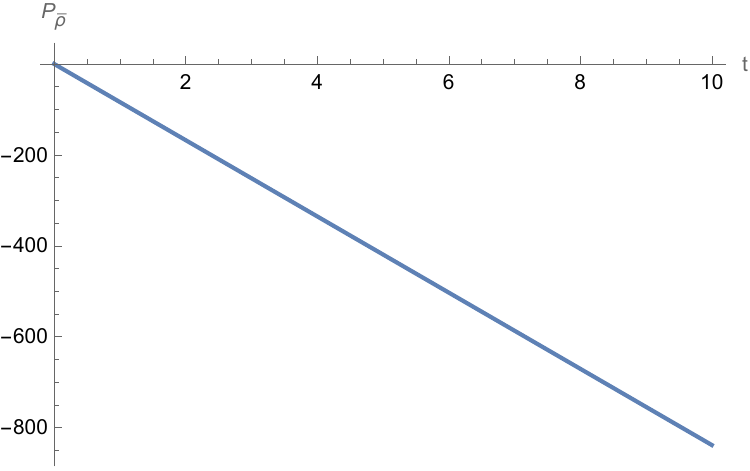}
     \includegraphics[width=0.45\linewidth]{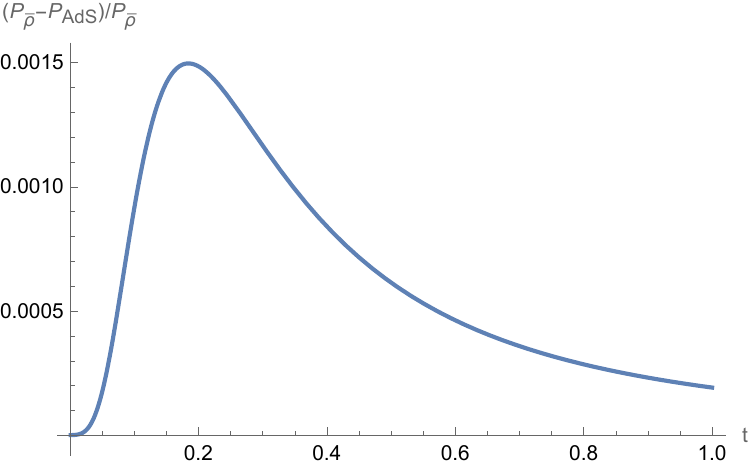}
    \caption{Proper momentum $P_{\bar \rho}$ and its comparison with pure AdS case for a smeared quiver with an $AdS_3$ factor in Example IV. We set $H_0=100$, $\eta_0=4$ and $P=50$.}
    \label{fig:Pex4}
\end{figure}

Next, we evaluate the integral \eqref{eta-t} for AdS$_3$ and AdS$_2$ cases separately.

 \underline{\textbf{For} $\mathbf{AdS_3}$}:
\begin{align}
\label{e3.35}
    \int^{\eta_{max}}_{\eta_0} \frac{d \eta}{A(\eta)\sqrt{A(\eta_0) e^{-\frac{\Phi(\eta_0)}{2}}-A(\eta)e^{-\frac{\Phi(\eta)}{2}}}}=\int_{\eta_{0}}^{\eta_{max}} d\eta\frac{\sin \left(\frac{\pi  \eta }{P+1}\right)\sin ^{1/8}\left(\frac{\pi  \eta_0 }{P+1}\right)}{\sqrt{1-\frac{\sin ^{1/4}\left(\frac{\pi  \eta_0 }{P+1}\right)}{\sin ^{1/4}\left(\frac{\pi  \eta }{P+1}\right)}}}.
\end{align}

\underline{\textbf{ For} $\mathbf{AdS_2}$}:
\begin{align}
\label{e3.36}
   \int_{\eta_{0}}^{\eta_{max}} \frac{d \eta}{A(\eta)\sqrt{A(\eta_0) e^{-\frac{\Phi(\eta_0)}{2}}-A(\eta)e^{-\frac{\Phi(\eta)}{2}}}}=\int_{\eta_{0}}^{\eta_{max}} d\eta\frac{\sin \left(\frac{\pi  \eta }{P+1}\right)\sin ^{1/4}\left(\frac{\pi  \eta_0 }{P+1}\right)}{\sqrt{1-\frac{\sin ^{1/2}\left(\frac{\pi  \eta_0 }{P+1}\right)}{\sin ^{1/2}\left(\frac{\pi  \eta }{P+1}\right)}}}.
\end{align}
The integrals in eqs. \eqref{e3.35}-\eqref{e3.36} can be evaluated numerically, or instead one can solve the differential equation 
\begin{align}
\label{e4.24}
    \frac{H_0^2 \dot{\eta}^2}{16}= \frac{  \left[A(\eta_0) e^{-\frac{\Phi (\eta_0)}{2}}-A(\eta)e^{-\frac{\Phi (\eta)}{2}}  \right]}{\sin ^{2}\left(\frac{\pi  \eta }{P+1}\right)\Big( \lambda^2 t^2 +4c_1\Big)^2}.
\end{align}
Notice that in order for the right-hand side ~of eq.~\eqref{e4.24} to be positive definite, one must have $\eta (t)> \eta_0$, which yields the trajectory along the $\eta$- coordinate, shown in Figure \ref{figsmeared}. In this case,  the particle moves along the direction of the increasing values of the rank function. 

In Figure \ref{fig:Pex4}, the proper momentum ($P_{\bar{\rho}}$) is plotted  
which at late times has a qualitatively similar behaviour as in the case with Poincare AdS. This is reflected in the fact that the momentum of the particle along the quiver direction freezes rapidly with time and at late times its dynamics is completely characterised by the motion in AdS.

To summarise, in all four examples above, we observe a non-trivial dynamics of the probe particle along the quiver (the $\eta$-axis), either in the increasing or the decreasing direction of the values rank function. In all these cases, the motion of the particle is damped in the $\eta$ direction, and as time progresses, the dynamics is fully driven by the Poincare AdS. 

\section{Conclusions and Future Work}\label{sectionconcl}

In this work we have provided the first systematic study of holographic
Krylov complexity in fully top-down AdS$_3$ and AdS$_2$
backgrounds dual to linear quiver CFTs and conformal quantum-mechanical
models. Unlike effective bottom-up AdS constructions, these geometries
possess a non-trivial dependence on the `quiver field theory coordinate'
($\eta$)  reflected in the rank functions $h_4,h_8$ that encode colour and flavour
structure in the dual field theories.
As a result, the motion of a massive probe
particle--and hence the holographic dual of operator growth--necessarily
involves simultaneous evolution in the radial direction $r(t)$
and along the `quiver field theory direction' $\eta(t)$. We refer to our calculation as Krylov complexity,  as it is a natural extension of that of \cite{Caputa:2024sux} for the case of warp factors and quiver structures.

We showed that this produces distinctive corrections to the
early-time behavior of the Krylov complexity, while at late times the
$\eta$-motion is strongly damped and the evolution universally approaches
that
of pure Poincaré AdS.

By studying a variety of representative quiver
solutions, including Abelian and non-Abelian T-dual backgrounds, 
quivers with localized flavor groups, and quivers with smeared flavors, we
demonstrated how quiver data quantitatively shapes the early-time
complexity growth through the geometry-operator-spreading correspondence
\cite{Caputa:2024sux}.

Our analysis reveals that holographic Krylov complexity offers a refined
probe of quiver dynamics, capturing features  beyond what can be
inferred from purely radial in-fall. The direction and magnitude of the
$\eta$-motion depend sensitively on the local slope of the rank functions
and on the quiver parameters
$(P, \mu,\nu,u_0)$, providing a dynamical holographic
mechanism for how operator growth
spreads across gauge nodes. The universality of the late-time regime
underscores an appealing physical picture: while early-time complexity
retains memory of UV quiver data, the system flows toward an IR regime
where operator growth effectively averages over (erases!) the quiver
structure. The
proper momentum thus encodes both local quiver geometry and emergent
universal CFT behavior.

It is worth pointing out here that, while the general connection 
between the rate of complexity growth and radial momentum was already 
conjectured \cite{Susskind:2018tei,Brown:2018kvn} in the case of AdS 
and AdS black holes, and argued from using the complexity-volume 
conjecture in \cite{Susskind:2020gnl} 
and in \cite{Barbon:2020uux} and previous work by the same authors, 
the precise relation (\ref{compmom}) for {\em Krylov complexity}, 
defined and checked in the 
case of $AdS_3$ 
in global and Poincar\'{e} coordinates and the $AdS_3$ black hole
in \cite{Caputa:2024sux}, 
is on a less strong footing. Thus, we have to 
understand the extension we proposed here to a more general warped 
AdS case as a conjecture, based on the understanding of 
holographic complexity in general. However, the general 
definition of Krylov complexity in field theory, and in particular 
in our case, of a Hilbert space containing a quiver space, as well 
as the regular CFT space, is still lacking, so a specific test of 
the conjecture is hard to do. A better chance of that stands 
the case of $AdS_2/CFT_1$, as then we have a quantum mechanical 
(as opposed to field theoretic) CFT, so we have a countable (and not 
continuous) Hilbert space, which we can thus order as $|0\rangle, 
|1\rangle,|2\rangle,...$, to which in principle we can apply the 
Lanczos algorithm and Gram-Schmidt orthogonalization with a 
(non-degenerate!) reference state $|0\rangle$ 
in order to obtain the Krylov basis, 
and thus compute the Krylov complexity. Another possibility would 
be the case that the warped AdS (or a similar gravity dual)
is considered in global coordinates,
corresponding to a CFT on a sphere, on which we can KK reduce, again 
obtaining a countable quantum mechanical system. These
calculations, possible in principle, we leave to further work.

The results in this paper open several interesting directions. First, it would be
natural to extend the present analysis to quiver gauge theories in three
and four dimensions. In analogy with recent work on ${\cal N}=4$ SYM
\cite{Fatemiabhari:2025cyy}, one could test whether multi-dimensional quiver
directions
produce additional channels of operator growth, and whether new forms of
damping or universality arise in higher-dimensional holographic flows,
for example in the systems of \cite{Akhond:2021ffz, Lin:2004nb, Gaiotto:2009gz, Nunez:2018qcj, Nunez:2019gbg, DHoker:2016ysh, Legramandi:2021aqv, Cremonesi:2015bld, Nunez:2018ags, Apruzzi:2013yva, Apruzzi:2015wna}, etc.
Second, it would be valuable to explore backgrounds that combine
non-trivial quiver warping with confinement, for example \cite{Chatzis:2024kdu, Chatzis:2024top, Fatemiabhari:2024aua, Chatzis:2025dnu, Chatzis:2025hek}, etc. Such geometries could clarify
how Krylov complexity interpolates between quiver-dominated UV behavior
and confining IR phases, potentially offering a new diagnostic of
confinement-induced slowdowns or plateaus in operator growth.
A further project along the same vein is to examine holographic setups where the 
probe motion is intrinsically multi-dimensional, requiring evolution not
only along $r$ and $\eta$
but also along an additional internal coordinate. This would provide
the first test of Krylov complexity in holographic
systems where operator growth explores a genuinely curved,
multi-coordinate internal space.
Such investigations could reveal
 novel qualitative spreading patterns and deepen the geometric
understanding of complexity in strongly coupled quantum systems.

Overall,
our results suggest that top-down holography provides a fertile arena for
uncovering the rich dynamical structure of Krylov complexity and its
dependence on microscopic field-theory data.

\section*{Acknowledgments}

For discussions, for comments on the manuscript, and for sharing their ideas with us, we wish to thank:  Dmitry Ageev, Nicol\'o Bragagnolo, Yolanda Lozano, Niall Macpherson, Alfonso Ramallo, Anayeli Ramirez, Ricardo Terrazas. 
The work of HN is supported in part by  CNPq 
grant 304583/2023-5 and FAPESP grant 2019/21281-4.
HN would also like to thank the ICTP-SAIFR for 
their support through FAPESP grant 2021/14335-0. C. N. is supported by 
STFC’s grants ST/Y509644-1, 
ST/X000648/1 and ST/T000813/1.
DR would like to acknowledge the Mathematical Research Impact Centric 
Support (MATRICS) grant (MTR/2023/000005) received from ANRF, India.

\bibliographystyle{JHEP}
\bibliography{main.bib}

\end{document}